# Methods for characterization of atomic-scale field emission point-electron-source

Shuai Tang[1], Mingkai Gou[1], Yingzhou Hu[1], Jie Tang[2], Yan Shen[1], Yu Zhang[1], Lu-Chang Qin[2], Ningsheng Xu[1], Richard G. Forbes[3], Shaozhi Deng[1]*

[1]State Key Laboratory of Optoelectronic Materials and Technologies, Guangdong Province Key Laboratory of Display Material and Technology, School of Electronics and Information Technology, Sun Yat-sen University, Guangzhou 510275, China

[2]Faculty of Materials and Manufacturing, Beijing University of Technology, Beijing 100124, China

[3]School of Mathematics and Physics, University of Surrey, Guildford, Surrey GU2 7XH, United Kingdom

Corresponding Author E-mail: stsdsz@mail.sysu.edu.cn

Abstract

Field emission (FE) electron sources are made close to atomic-scale to reach the highest spatial resolution as well as stable emission for electron microscopy, electron beam inspection and lithography. At present, no single agreed method exists of using FE current-voltage data to extract the apparent emission area, which is needed for predicting some beam properties. The 1956 theory of Murphy and Good (MG) is better physics than the 1920s theory of Fowler and Nordheim (FN) and colleagues, but many researchers use simplified FN theory to analyse experimental data. The present paper reports an experimental method of finding apparent emission area, based on using field ion and field electron microscopes (FIM-FEM). The discrepancy of emission area between the FIM-FEM method and MG-based analysis is a factor of 7.4, while that with simplified FN-based analysis is about 25, confirming MG theory is better for FE data analysis. The result allows deduction of key indicators, including source energy spread, reduced brightness and emission efficiency. A downloadable program is made available to help analysis. Our work provide a new experimental method of characterizing FE electron sources, especially the atomic-scale cold cathode, for which existing plot-based data-analysis methods are not suitable.

## 1. Background

Electron microscopy has a wide range of applications in integrated circuits, materials science, structural biology, and ultrafast science. With the miniaturization of chip structures [1, 2] and the development of cryo-electron microscopy [3, 4], ultra-fast electron microscopy [5], in situ liquid electron microscopy [6] and other new technologies, electron microscopes need extremely high spatial resolution at the picometer level [7, 8] and temporal resolution at the attosecond level [9]. The spatial resolution of an electron microscope is determined by the virtual source characteristics, the spherical aberration of the electromagnetic lens, chromatic aberration of the electron beam, and diffraction of the electron beam [10–12]. Since spherical aberration issues have been resolved [13, 14] and diffraction is determined by the accelerating voltage, the characteristics of the electron source (including size and energy spread) become the key factors limiting the spatial resolution of the electron microscope [15]. Emission induced by a highly negative electrostatic (ES) field is called here "field electron emission" (FE). Compared with other cathode types, FE sources offer the highest brightness and the narrowest total-energy spread. These things bring benefits for high-resolution imaging and spectroscopy. Since the pioneering FE theory reported in 1928 [16], various materials including metals [17, 18], metal oxides [19–21], carbon-based nanostructures [22–24], metal carbides [25–27] and rare-earth hexaborides [28–30] have been investigated and used as cold cathodes.

The lanthanum hexaboride ($LaB_6$) cold cathode, with apex radius around 10 nm, exhibits outstanding performance. It has better signal-to-noise characteristics than a tungsten cold cathode, and good resolution inside a SEM [31]. Our recently developed [28] very-fine-tip $LaB_6$-nanocone cold cathode, fabricated by a FIB-based method, performs better than commercial tungsten sources when operated in a spherical-aberration-corrected TEM [28, 32]. This $LaB_6$ FE source exhibits equivalent high spatial resolution, one third lower energy spread, and more than tenfold enhanced stability of tungsten FE source.

Much of this excellent source behaviour can be attributed to its small apex radius of 10 nm. This results in a smaller virtual source size as well as more stable emission characteristics. Particularly with the growing use of small-apex-radius emitters, it is very desirable to establish theoretical and experimental methodologies for more accurate analysis of the performance of FE sources in general.

FE theory involves quantum tunnelling. The 1928 Fowler-Nordheim (FN) paper [16] evaluated tunnelling probabilities by using an exactly triangular (ET) tunnelling barrier. This approach ignored image-potential-energy (PE) effects. This incorrect approach is also used when deriving the elementary FE equation, which often appears in technological FE papers and is written here as

$$I_{\mathrm{m}}^{\mathrm{EL}} = \frac{A_{\mathrm{f}}^{\mathrm{ET}} a \beta^2 V_{\mathrm{m}}^2}{\phi} \exp\left(\frac{-b\phi^{\frac{3}{2}}}{\beta V_{\mathrm{m}}}\right), \tag{1}$$

where $a$ [$\approx 1.541\times10^{-6}$ A eV V$^{-2}$] and $b \approx 6.831\times10^{9}$ eV$^{-3/2}$ V m$^{-1}$] are the FN universal constants [see electronic supplementary material (ESM)], and $A_{\mathrm{f}}^{\mathrm{ET}}$ is the formal emission area derived by assuming an ET barrier. The parameter $\beta$ that links the magnitude $F_{\mathrm{a}}$ of the electrostatic field at the emitter apex to $V_{\mathrm{m}}$ is defined via

$$F_{\mathrm{a}} = \beta V_{\mathrm{m}}. \tag{2}$$

$\beta$ is called here the (voltage-to-apex-field) *conversion factor (CF)*. For so-called *electronically ideal* systems $\beta$ can be treated as constant (see ESM for details).

Later in 1928 [33], Nordheim realized that including image PE effects, to give what is now called the Schottky-Nordheim (SN) tunnelling barrier (see ESM), leads to a correction factor in the exponent of eq. (1). Unfortunately, he made a mathematical mistake [34]. In the 1950s [35], Burgess, Kroemer and Houston (BKH) suggested that the correction factor should be a particular value of a mathematical function denoted here by $\mathrm{v}_{\mathrm{BKH}}(y)$, where $y$ is the *Nordheim parameter* (see ESM). Murphy and Good [36] then developed a replacement for the 1928 FN FE equation. This replacement is called here the 1956 Murphy-Good (MG) FE equation.

A modern development is the so-called "Extended" (or "Experiment-oriented") MG (EMG) FE equation:

$$I_{\mathrm{m}}^{\mathrm{EMG}} = \frac{A_{\mathrm{f}}^{\mathrm{SN}} a \beta^2 V_{\mathrm{m}}^2}{\phi} \exp\left(\frac{-\mathrm{v}_{\mathrm{FD}}(f)\cdot b\phi^{\frac{3}{2}}}{\beta V_{\mathrm{m}}}\right). \tag{3}$$

Here $I_{\mathrm{m}}^{\mathrm{EMG}}$ is the measured current (as described by EMG theory) and $A_{\mathrm{f}}^{\mathrm{SN}}$ is the formal emission area derived by assuming a SN barrier. The parameter $f$ is the so-called *scaled field* (see ESM), and $\mathrm{v}_{\mathrm{FD}}(f)$ is a particular value of a special mathematical function introduced by Forbes and Deane (see [37] and ESM). Because it is close to unity, the factor $\mathrm{t}^{-2}$ in the 1956 MG FE zero-temperature equation has been absorbed into $A_{\mathrm{f}}^{\mathrm{SN}}$.

Both eq. (1) and eq. (3) lead to area extraction procedures, at present usually based on FN plots. However, the formal area extracted by assuming a SN barrier is much less (by a factor of 100 or more) than the formal area extracted by assuming an ET barrier. Thus, it is of practical interest which barrier is the better approximation to reality, since this affects the numerical values deduced from current-voltage experiments for current densities, brightness and other key indicators of a field emitting electron source. It is theoretically certain that assuming a SN barrier is better physics than assuming an ET barrier, and that MG-type FE theory is better physics than 1920s FE theory [34]. However, for scientific completeness it would be useful to also have experimental proofs of this.

A useful approach was developed in 1973 by Ehrlich and Plummer [38], who attempted direct measurement of the local emission current density (LECD) from a tungsten (110) plane. They found, as theoretically expected, that the experimental results were significantly closer to MG FE theory than to elementary FE theory. Their conclusion seems qualitatively valid, but interpretation of their results is not without its difficulties and the emission area cannot be accurately obtained from their approach.

This paper aims to provide a new experimental method for measuring the source area of a single needle-like electron source. This new method is expected to work both for relatively blunt emitters and for sharp emitters for which the planar emission approximation (used in deriving emission current formulas—see ESM) seriously breaks down. Thus, our method will work for the upcoming atomic-scale cold cathode. Further, for "not too sharp" emitters, the new method provides further experimental evidence that MG FE theory is more accurate than elementary FE theory.

## 2. Experimental and theoretical methods

Our new approach is based on the techniques of field ion microscopy (FIM) and field electron microscopy (FEM) and related theory, and involves investigations to determine the linear and area magnifications that apply to these techniques, in particular to our specific experimental system. Our thinking is that, if the area magnification can be successfully determined, then the total area of a FEM screen image can be converted back to an electron-source area on the emitter, and that this constitutes an alternative method of extracting emission area.

It is simple and convenient to calculate linear and area magnifications when using a sharply pointed electron source. A sharp source prepared from a HfC nanowire has been chosen for these FIM-FEM measurements.

### 2.1 Simulation methods for obtaining the particle trajectories in FIM and FEM

An initial requirement is using finite-element calculations to confirm that (a) electrons in FEM and (b) hydrogen ($H_2^+$) ions in FIM follow essentially the same trajectories if both start with zero kinetic energy, and hence that the two microscopies have the same transverse and area magnifications.

Our simulations used the following geometry, based on our experimental system. The emitter is modelled as a narrow-angle cone with a rounded apex of radius from 5 nm to 25 nm. The spacing between the emitter and an extractor ring is 2 mm; the ring diameter is 10 mm. The extractor voltage is set to 0 V. A microchannel-plate detector is at a distance 50 mm from the emitter apex, with its beam-facing surface also at 0 V. The electron/ion source radius ($R_{\mathrm{emission}}$) at the emitter tip is set from 0.5 nm to 2 nm. Several particle-monitoring surfaces along the system axis record particle behaviour at different positions along their trajectories. The distribution area, velocity, angle relative to the axis and other physical quantities were thus monitored at different distances from emitter apex. Simulations used the CST Studio Suite. In the FIM simulation, using $H_2^+$ ions, the emitter voltage is set as +2000 V; the electrostatic (ES) field at the emitter apex is found as about 37 V/nm. In the FEM simulation, using electrons, the emitter voltage is set as –150 V; the apex ES field is found as –2.76 V/nm.

## 2.2 FIM and FEM experimental methods

The common emitter used in both FIM and FEM was prepared, by electropolishing and then in-situ field evaporation in FIM mode, from a [100]-oriented single-crystal HfC-nanowire emitter with a square cross-section of side-length 80 nm. The resulting apex radius was approximately 25 nm. The FIM experiments were carried out at room temperature with a background pressure of $1.0\times10^{-7}$ Pa. This situation is not ideal, but is what our current experimental system allows. The operating-gas ($H_2$) pressure was about $6.5\times10^{-4}$ Pa. The emitter voltage was between +2000 V and +5000 V. Operating-gas molecules are field-ionized close to the emitter surface. Resulting ions are guided by the field to hit the microchannel-plate: this generates a FIM image on a phosphor screen, as shown in Fig. 1(a).

The FEM experiments were conducted in the same apparatus, using a high vacuum of $1.0\times10^{-7}$ Pa and with emitter voltages between –400 V and –1000 V. Field emitted electrons produce at the microscope screen an image of the electron source of area $A^{\mathrm{S}}$ on the emitter tip, as shown in Fig. 1(b). Comparing the results of the FEM and FIM experiments, taking into account the known crystallography of the emitter apex, enables an estimate of $A^{\mathrm{S}}$, as described below.

## 2.3 The FN-plot method

Discussion here uses the SI unit "volt" (V) for measured voltage $V_{\mathrm{m}}$ and the SI unit "ampere" for measured current $I_{\mathrm{m}}$. Related data plots are made in the " *FN coordinates*" $\ln\{I_{\mathrm{m}}/V_{\mathrm{m}}^2\}$ vs $V_{\mathrm{m}}^{-1}$.

MG theory predicts FN plots to be slightly curved [35, 39], more so at the left-hand (high-voltage) side. This creates errors if plot analysis uses ET-barrier theory, but mathematical complications when SN-barrier theory is used. The theory needed for the SN-barrier case is discussed elsewhere (Appendix to [40]), and set out in the ESM. A summary follows.

Let the straight line fitted to an experimental FN plot have the form

$$L^{\mathrm{expt}}(V_{\mathrm{m}}^{-1}) \equiv \ln\{I_{\mathrm{m}}/V_{\mathrm{m}}^2\} = \ln\{R^{\mathrm{expt}}\} + S^{\mathrm{expt}}/V_{\mathrm{m}}\,, \tag{4}$$

where $R^{\mathrm{expt}}$ and $S^{\mathrm{expt}}$ are parameters expressed in A/V$^2$ and Np V, respectively, with $S^{\mathrm{expt}}$ negative.

A validity check, such as the orthodoxy test [41], should be used to investigate whether the related FE system is electronically ideal. If this is passed (or if the results are indecisive but are close to the pass range), then values of $R^{expt}$, $S^{expt}$, and the combination $[R^{expt} \cdot (S^{expt})^2]$ are recorded, as in Table 1 below.

To implement data analysis, this paper uses the parameter "scaled field" ($f$ above) and the related modern version of the tangent method (see ESM).

Strictly, the fitted line $L^{expt}(V_m^{-1})$ is a chord to the data points [42]. The tangent method models this chord by a tangent to the chosen theoretical equation when this theoretical equation is formulated and plotted in FN coordinates. This approach is a good approximation.

The theoretical tangent must be taken at the value of $V_m^{-1}$ at which this tangent is parallel to the experimental chord. This $V_m$-value is called the *fitting value* and denoted by $V_t$. Corresponding to $V_t$ there are fitting values $F_t$ of apex "field" and $f_t$ of the related scaled field. Best estimates of $V_t$, $F_t$ and $f_t$ are initially undecided.

FN-plot analysis involves three correction functions defined in the ESM: the function $v_{FD}(f)$ discussed earlier; the slope correction function $s(f)$; and the intercept correction function $r(\phi, f)$. For simplicity, the fitting values of $s(f)$ and $r(\phi, f)$ are denoted by $s_t$ and $r_t$, respectively.

Extracted values of the conversion factor $\beta$ and of the formal emission area $A_f^{SN}$ that appear in eq. (3) are given by (see ESM):

$$\beta^{extr} = -s_t \, b\phi^{3/2} / S^{expt}, \tag{5}$$

$$\{A_f^{SN}\}^{extr} = \Lambda_{FN}^{SN} \cdot [R^{expt} (S^{expt})^2], \tag{6}$$

where the *area extraction parameter*, $\Lambda_{FN}^{SN}$, for a FN plot, assuming a SN barrier, is

$$\Lambda_{FN}^{SN} = 1/[(ab^2\phi^2) \cdot (r_t s_t^2)]. \tag{7}$$

Formulae for FN-plot analysis using the ET barrier are obtained by setting $r_t$=1 and $s_t$=1 (see ESM).

When SN-barrier theory is used, a procedure is needed for determining "best estimates" for $s_t$ and $r_t$. The procedure used in this paper is described in Section 4.2 and the ESM.

## 3. Results and discussion

### 3.1 Simulation results for magnification ratio between FIM and FEM

Figure 1(a) shows how the radius of the emitted electron or ion beam (i.e., the radius of its envelope) varies with distance along the system axis, measured from the emitter surface. Red dots show $H_2^+$ data and green dots electron data: the two lines of dots overlap perfectly. The blue circles show the ratio of FEM-results/FIM-results: this ratio has a constant value of 1.00. Thus, these simulations show that electrons and $H_2^+$ ions emitted with zero kinetic energy from a given point on the emitter surface follow the same trajectory. Historically, this has been a widely accepted result, but it is difficult to find any formal theoretical proof.

In a FIM or FEM, a distance $\rho_{scr}$ on the microscope screen corresponds to a distance $\rho_{em}$ on the emitter surface. The microscope's *linear magnification* $m_{lin}$ is defined as $\rho_{scr}/\rho_{em}$. The simplest assumption, made here, is that $m_{lin}$ has the same constant value for all parts of the image and all directions in the image. The microscope's area magnification $M$ is then given by $M = (m_{lin})^2$.

For the microscope geometry and assumed emitter shape described above, we investigated how the area magnification $M$ depends on the apex radius-of-curvature and on the radius $R_{emission}$ of the source area. Results are shown in Fig. 1(b). As expected, the area magnification is mainly affected by the apex radius-of-curvature. The source radius does not significantly affect the area magnification. For our experimental emitter of apex radius about 25 nm, the linear magnification predicted by these simulations is about $1.4\times10^6$.

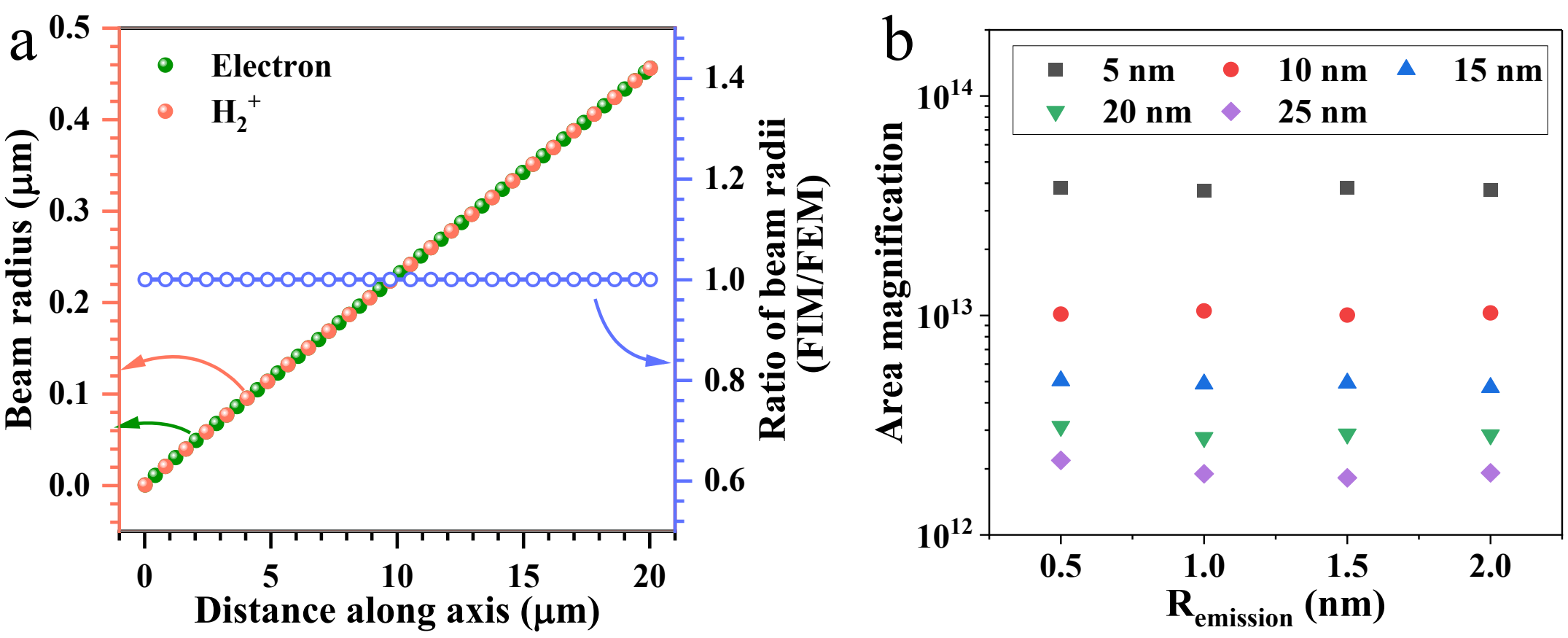

Fig. 1. Data relating to the charged-particle optics of field electron and field ion microscopes. (a) To show how ion ($H_2^+$) and electron beam radii vary with distance along the system axis (results derived by finite-element simulations). (b) To show how the area magnification varies with the radius $R_{emission}$ of the emission area, for different values of the apex-radius-of-curvature (top figures).

### 3.2 Deriving electron source area by using FIM&FEM charged-particle optics

Since the FEM area magnification $M_{FEM}$ has been proven the same as the FIM area magnification $M_{FIM}$, the electron source area $A^S$ on the emitter can be found by using the formula $A^S = A^{FEM}/M_{FIM}$, where $A^{FEM}$ is the area of the FEM-mode image spot on the microscope screen. Fig. 2(b) shows a typical FEM image for the HfC nanowire with an oxycarbide surface. $M_{FIM}$ and $A^{FEM}$ are found as follows.

After preparation for microscopy, the nanowire has a relatively sharp tip, with apex radius about 25 nm. HfC has a "rock-salt" crystal structure, which is a form of face-centred-cubic (fcc) structure, and has a lattice constant $c_{\mathrm{HfC}} = 0.456$ nm. The central lattice plane in the FIM image (Fig. 2(a)) is the 001 plane. In this plane the nearest-neighbour separation is $c_{\mathrm{HfC}}/\sqrt{2} = 0.322$ nm [43].

Although image quality is limited, we consider that we can identify two adjacent atoms in the net plane. In Fig. 2(a) the centres of the corresponding image spots are joined by a red line. On the photographic record the length of this line is 1.7 mm. The actual separation of spot centres on the microscope screen is determined as follows.

The channel plate has a small hole of known diameter 2.0 mm. We assume that its circumference is, in effect, imaged directly onto the microscope screen as a 2 mm diameter circle. It thus appears in our photographic image record, where its diameter is measured as 4.5 mm. Assuming magnification can be treated as uniform across the image, it follows that distance $\rho_{\mathrm{rec}}$ on the image record is converted to distance $\rho_{\mathrm{FIM}}$ on the microscope screen by $\rho_{\mathrm{FIM}} = \rho_{\mathrm{rec}}/2.25$. This yields the actual separation of the adjacent image-spot centers on the microscope screen as 0.76 mm.

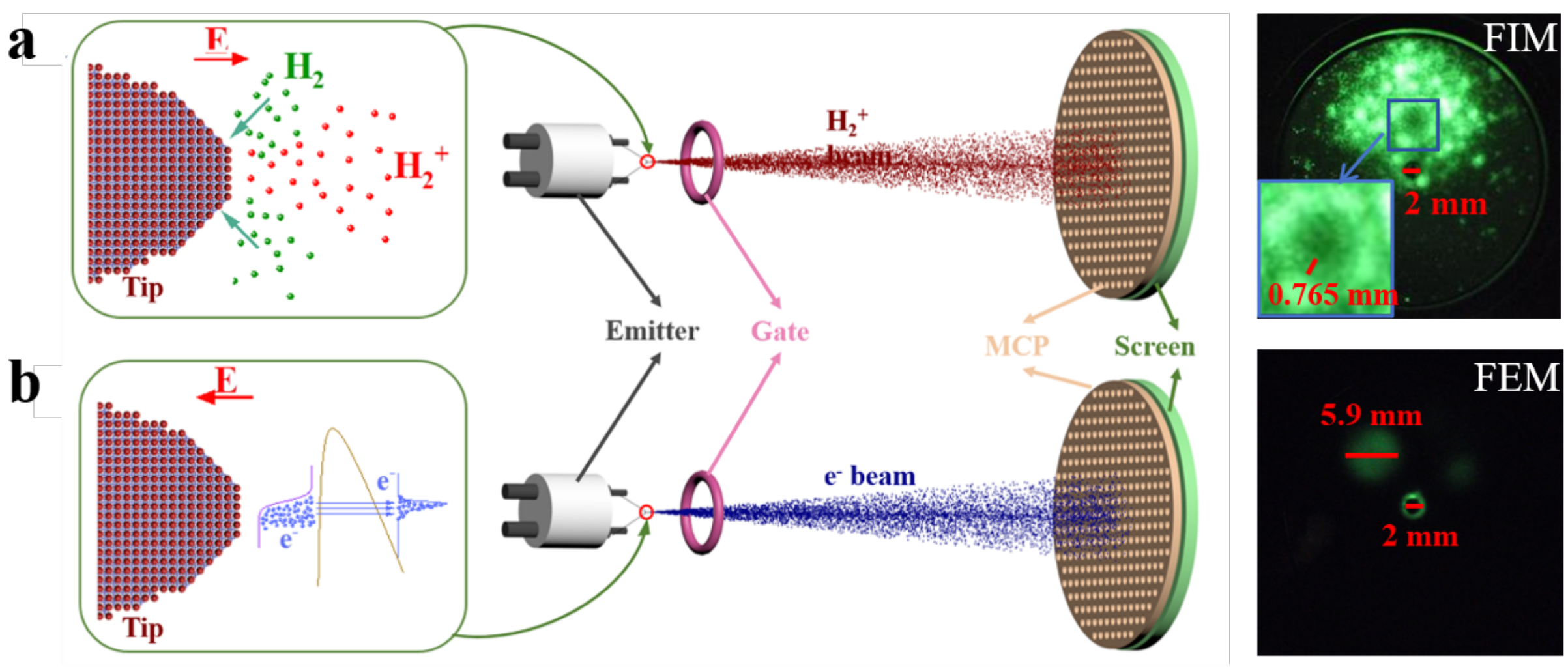

Fig. 2. (a) Schematic diagram of FIM operation, and FIM image of HfC emitter. (b) Schematic diagram of FEM operation, and FEM image of same HfC emitter.

In the (001) net plane, the nearest-neighbour distance (both between Hf atoms and between C atoms) is $c_{\mathrm{HfC}}/\sqrt{2} = 0.322$ nm. As shown in Table 1, this yields, for the geometry of our microscope and HfC specimen, a linear magnification of about $2.4\times10^{6}$. For comparison, the simulation result was about $1.4\times10^{6}$. Given our limited knowledge of the experimental emitter shape, this degree of agreement is considered very satisfactory.

The field emitted electrons produce at the microscope screen an image of the electron source area on the emitter tip, as shown in Fig. 2(b). An initial estimate of the physical radius of the screen image can be found, as above, by making use of the known dimensions of the channel-plate hole, and is recorded in Table 1.

However, we also need to take into account the possible influence of beam effects that would tend to expand the image laterally. These include (e.g., [44], p. 1005) expansion due to mean space-charge effects and expansion associated with the Boersch spatial effect which results from stochastic coulomb interactions. Due to current-density differences, these effects could be significant for electrons but not for $H_2^+$ ions. If a broadening effect occurs, it would be necessary to divide the initial spot-estimate deduced above by a correction factor related to this effect.

Table 1: Data relating to FIM and FEM experiments. (For precision in calculations, intermediate values are shown to 3 significant figures. Accuracy of final result is not better than 10%.)

| Line | Parameter | Numerical value for HfC |
|---|---|---|
| 1 | Lattice constant (nm) | 0.456 |
| 2 | Orientation of central facet | (001) |
| 3 | Nearest-neighbour separation in central facet (nm) | 0.322 |
| 4 | Separation of FIM image-spot centres on FIM screen (mm) | 0.76 |
| 5 | Deduced linear magnification of FIM | $2.36\times10^6$ |
| 6 | FEM image-spot radius on FEM screen (with beam broadening) (mm) | 2.95 |
| 7 | Estimated spot-radius correction factor | 1/1.2 |
| 8 | Estimated FEM image-spot radius on FEM screen (without beam broadening) (mm) | 2.46 |
| 9 | Estimated radius $\rho^S$ of FE electron source on emitter (using FIM linear magnification) (nm) | 1.041 |
| 10 | Derived FE electron source area $A^S$ on emitter (nm$^2$) | 3.41 |

Making reliable theoretical estimates of these beam effects is difficult, so we use Fig. 2(f) in Fink's1988 paper [45] on point sources to make an empirical estimate of the (maximum) linear-magnification difference between electrons and ions. Fink's Fig. 2(f) is a FEM image of a 3-atom tungsten tip, with the corresponding FIM image superimposed in reverse contrast. Measurement on an electronically enlarged version of Fink's Fig. 2(f) suggests that the ratio of his FEM image radius to his FIM image radius is approximately 1.2. Some of this difference may be caused if the two imaging processes image slightly different areas on the tip, but this figure of 1.2 should provide a maximum estimate of the bean broadening correction factor as it applies in Fink's experiment. Lacking better information, we assume it applies to our experiments, and that the "FEM screen-image radius without broadening" is obtained by dividing the "initial FEM screen-- image radius" (derived above) by the correction factor 1.2. Numerical details are in Table 1.

This result is then converted to a linear distance $\rho^{S}$ on the emitter tip, using the linear magnification $m_{lin}$ derived from the FIM experiments. This distance $\rho^{S}$ is the apparent FE source radius on the tip (i.e. $\rho^{S}$ is the extracted experimental value of the parameter $R_{emission}$ used in the simulations). Finally, in line 10, $\rho^{S}$ is converted to a corresponding electron-source area $A^{S}$, by ignoring small complications due to emitter-tip curvature and using $A^{S}=\pi\{\rho^{S}\}^{2}$. The final result is $A^{S} \sim 3.4$ nm$^{2}$.

We think it possible that details of the electron optics of the linkage between the channel plate and the microscope screen could mean that the image-spot radius at the microchannel-plate entry plane is not exactly equal to the image-spot radius on the microscope phosphor screen. This would require a more careful analysis and a further correction factor. However, any such factor is expected to be very close to unity, and should not affect the final conclusions below.

### 3.3 Emission area derived by FN-plot analysis

Using the method of Section 2.3, the formal emission area of the HfC emitter has been extracted from a FN plot, using the alternative assumptions of (a) an ET barrier and (b) a SN barrier. Fig. 3(a) shows the current-voltage data collected, and Fig. 3(b) the related FN plot.

Results are shown in Table 2. Lines 2 to 8 report experimental data as derived from the relevant FN plot.

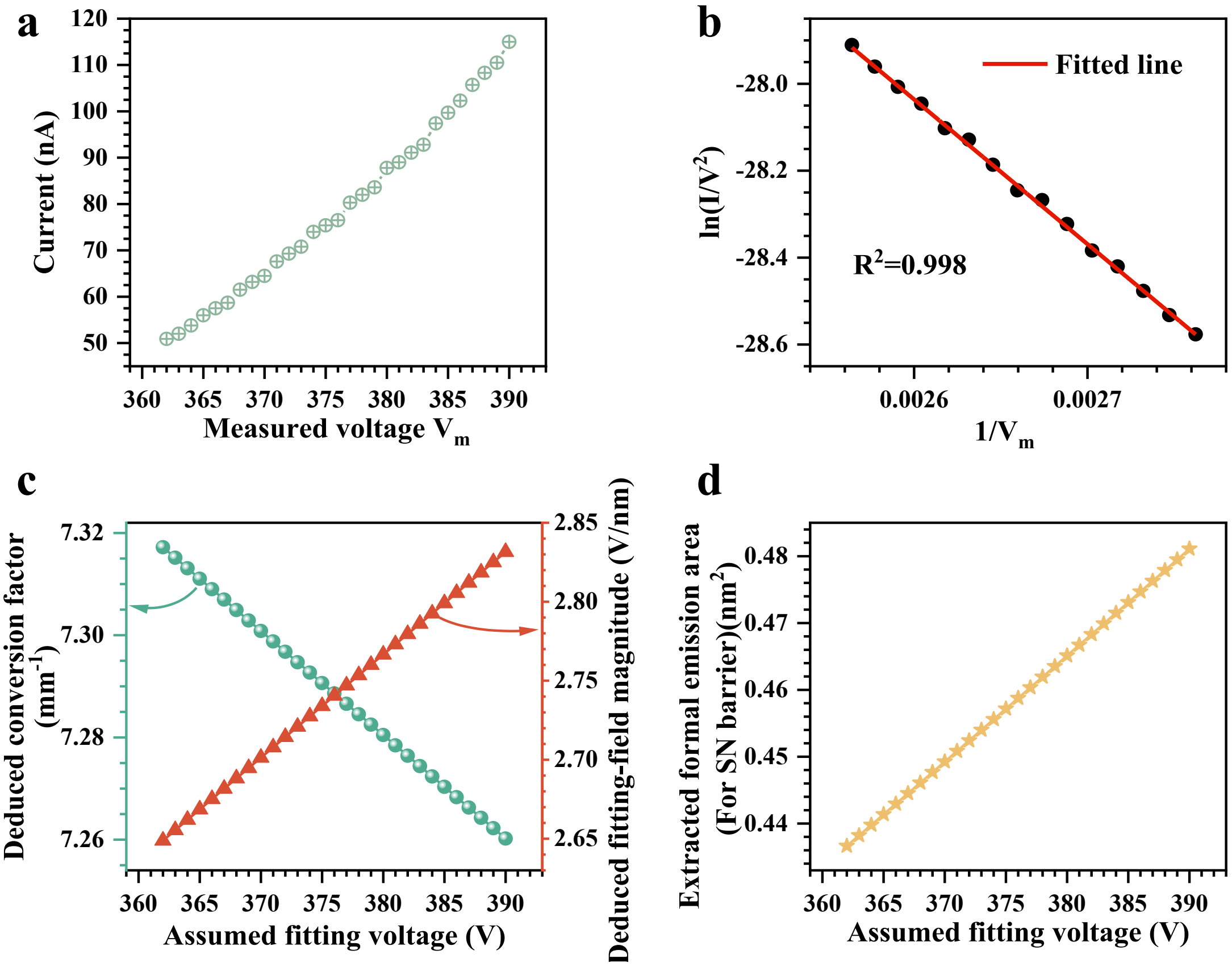

Fig. 3. (a) Current-voltage measurements for HfC emitter. Reproduced with permission from Tang et al., Nano Res. 13, 1620-1626 (2020). Copyright 2020 Springer Nature. [25] (b) Related FN plot. (c) To show how the extracted values of conversion factor and fitting field depend on the value chosen for the fitting voltage. (d) To show how the extracted value of formal emission $A_f^{SN}$ depends on the value chosen for the fitting voltage.

Results are shown in Table 2. Lines 2 to 8 report experimental data as derived from the relevant FN plot. As explained earlier, in the SN barrier case the extracted values of $\beta$ and $A_f^{SN}$ depend on what value is assumed for the "fitting value" $V_t$ of measured voltage: for a given fitting voltage, an iteration procedure (described in Section 4.2 and the ESM) is used to establish the corresponding values of $\beta$, $F_t$, $f_t$, $s_t$ and $r_t$. Line 9 in Table 2 reports the "starting value" $\beta_l$ used in these iterations.

Consequences of this need to decide $V_t$ are illustrated in Figs 3(c) and 3(d). The effect on the extracted value of $\beta$ is slight (the range is about 1%). However, due to the "lever effect" that occurs with semi-logarithmic plots, the effect on the extracted value of $A_f^{SN}$ is more significant (the range is about 10%).

As a convenient "definitive" approximation, we take $V_t$ to be the mid-range voltage $V_{mid}$ recorded on line 7 of Table 2. Lines 10 to 15 report the corresponding values of $\beta$, $F_t$, $f_t$, $s_t$, $r_t$, and $r_t s_t^2$. Lines 16 and 17 show the corresponding extracted values of $A_f^{ET}$ and $A_f^{SN}$.

The true fitting value of measured voltage will be slightly higher than $V_{mid}$, so using (instead) the voltage-range maximum will provide a maximum bound (here 6%) on the error associated with using the mid-range value. Details are provided on lines 18 and 19. This uncertainty is unavoidable when FN plots are used.

Table 2: Data relating to comparison of emission areas.

| Line | Parameter | Numerical value for HfC [25] |
|---|---|---|
| 1 | Assumed local work function $\phi$ (eV) | 2.50 |
| 2 | Vertical axis intercept ln{$R^{expt}$} | –19.38 |
| 3 | $R^{expt}$ (A/V$^2$) | $3.832\times10^{-9}$ |
| 4 | FN plot slope $S^{expt}$ (Np V) | –3329 |
| 5 | $R^{expt}\cdot(S^{expt})^2$ (A) | $4.246\times10^{-2}$ |
| 6 | Experimental voltage range (V) | 362–390 |
| 7 | Mid-range experimental voltage $V^{mid}$ (V) | 376 |
| 8 | Related experimental current $I$ (nA) | 75 |
| 9 | CF starting value $\beta_1$ (nm$^{-1}$) | $1.000\times10^{-3}$ |
| 10 | Extracted CF value $\beta$ (nm$^{-1}$) [assuming $V_{mid}$ as the fitting voltage] | $7.26\times10^{-3}$ |
| 11 | Fitting value $F_t$ of apex field (V/nm) | 2.73 |
| 12 | Fitting value $f_t$ of apex scaled field | 0.629 |
| 13 | Fitting value $s_t$ of slope correction factor | 0.892 |
| 14 | Fitting value $r_t$ of intercept correction factor | 282 |
| 15 | Value of $r_t s_t^2$ | 224 |
| 16 | Formal emission area extracted using ET barrier $A_f^{ET}$ (nm$^2$) | 90 |
| 17 | "Mid-range" formal emission area extracted using SN barrier $A_f^{SN}$ (nm$^2$) | 0.46 |
| 18 | Highest limit for $A_f^{SN}$ (nm$^2$) | 0.488 |
| 19 | Upper bound on extraction error for $A_f^{SN}$ | 6 % |
| | *Comparisons using present data* | |
| 20 | $A_f^{ET}/A^S$ | ~ 25 |
| 21 | $A^S/A_f^{SN}$ | ~ 7.4 |
| | *Corresponding Ehrlich & Plummer data* [38] | |
| 22 | $j_m/j_{EL}$ | ~ 20–40 |
| 23 | $j_{MG}/j_m$ | ~ 4.4–4.8 |

Lines 20 and 21 in Table 2 then use ratios to compare the emission area $A^S$ extracted from the FIM and FEM experiments with the results obtained from FN plots as analysed using ET and SN barriers. Clearly, the FIM/FEM experimental result is much closer to the SN barrier theoretical result. These new results are qualitatively consistent with the older results of Ehrlich and Plummer, shown in Lines 22 and 23.

## 4. General discussion

### 4.1 Some implications of accepting MG theory

There are various implications when experimentalists use EMG FE theory (rather than elementary FE theory) to analyse their experimental current-voltage data.

#### 4.1.1 Some implications for single-tip field emitters (STFEs)

As already indicated, extraction of voltage-to-local-field conversion factors ($\beta$-values) is only slightly affected by the use of EMG FE theory - extracted values are typically roughly around 5% higher when using elementary FE theory. However, using elementary theory leads to over-prediction of extracted formal area by a factor of order 100 or more and under-prediction of local emission current density and hence emission currents by a factor of 100 or more. Such errors could be highly unfortunate in research-and-development contexts – for example when trying to report the effectiveness of some new form of emitter, or to predict the occurrence of field-emission-initiated electrical breakdown. For professional engineers, use of theory known to be seriously incorrect might be considered irresponsible behaviour.

A context where MG FE theory has long been used is electron-optical theory. An important parameter [10] is the parameter $d_{\mathrm{F}}$ called here the "*D*-decay width at the Fermi level". This is defined and given (in MG FE theory) by

$$d_{\mathrm{F}} \equiv 1/\left[(-\partial \ln D/\partial H)|_{H=\phi}\right] = \mathrm{t}^{-1}\cdot(2/3b)(F/\phi^{1/2}) \quad , \tag{8}$$

where $D$ is the tunnelling probability for a SN barrier of zero-field height $H$, $b$ is the second FN constant as before, and t is the parameter that appears in the pre-exponential of the 1956 MG FE equation. The combination $(2/3b)$ is a universal constant with the value 9.760 $\mathrm{eV}^{3/2}\,(\mathrm{V/nm})^{-1}$, and t can usually be approximated (to the nearest 5%) as 1.05. The parameter $d_{\mathrm{F}}$ plays an important role in the theoretical prediction of energy distributions [10], in particular the room-temperature total-energy distribution (TED) [46] shown in Fig. 4(a).

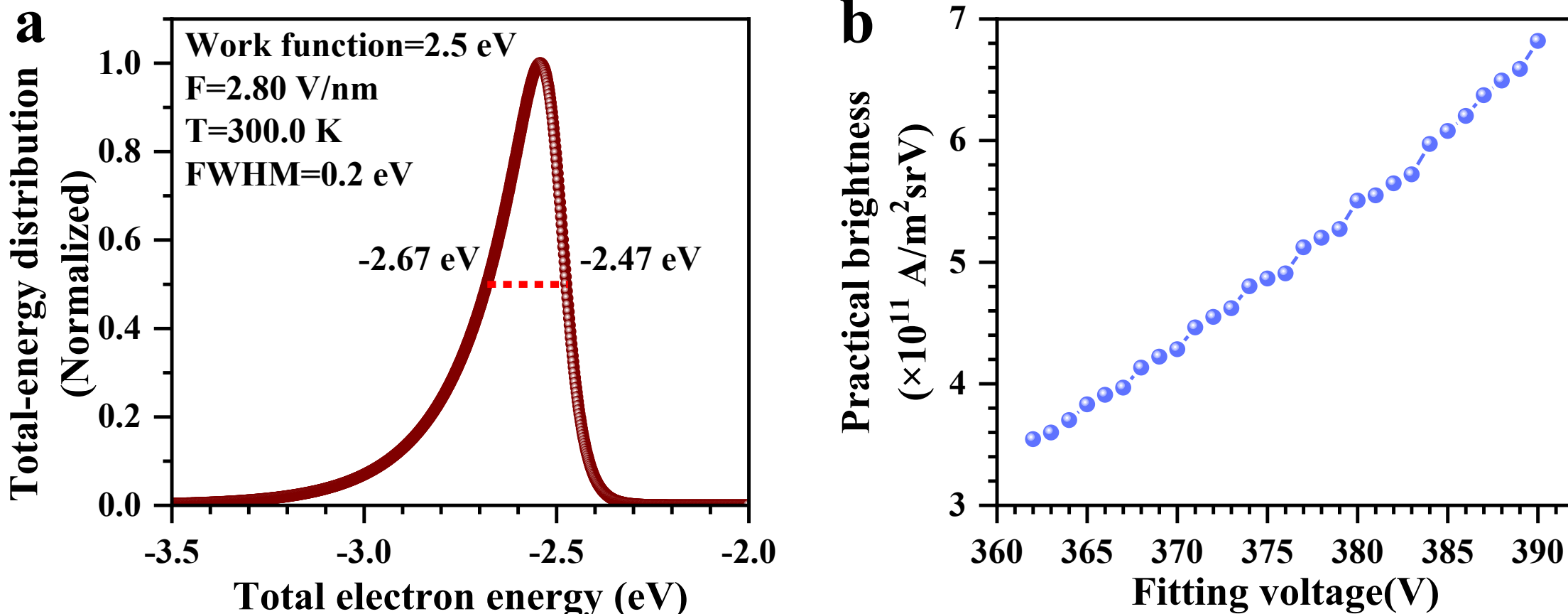

Fig. 4. (a) Predicted normalized room-temperature total-energy distribution (TED) for 2.5 eV work-function emitter, at surface field of 2.8 V/nm. (b) To show how the practical brightness $B_{pract}$ extracted using eq. (11) depends on the value chosen for the fitting voltage $V_t$.

Another parameter of electron-optical interest is the "practical brightness" $B_{pract}$ discussed in Ref. 11 and estimated in the present context by

$$B_{\mathrm{pract}} \approx 1.44\, J^{\mathrm{extr}}/\pi d_{\mathrm{F}} \approx (1.44/\pi d_{\mathrm{F}})(I_{\mathrm{m}}/A_{\mathrm{f}}^{\mathrm{SN}}) \ , \tag{9}$$

where $J^{\mathrm{extr}}$ $[\equiv I_{\mathrm{m}}/A_{\mathrm{f}}^{\mathrm{SN}}]$ is the "extracted LECD". Because both $A_{\mathrm{f}}^{\mathrm{SN}}$ and this extracted LECD value depend on the value assumed for the fitting voltage $V_t$, the value of $B_{\mathrm{pract}}$ derived from eq. (9) will also depend on the value assumed for $V_t$, as shown in Fig. 4(b).

4.1.2 Extraction of formal area efficiency

For FN plots, the extraction arguments above also apply to so-called large area field electron emitters (LAFEs), which contain very large numbers of individual emission sites.

For LAFEs, an additional parameter, namely the extracted *formal area efficiency* $\alpha_{\mathrm{f}}^{\mathrm{SN}}$ can be defined by

$$\alpha_{\mathrm{f}}^{\mathrm{SN}} \equiv A_{\mathrm{f}}^{\mathrm{SN}}/A_{\mathrm{M}} \ , \tag{10}$$

where $A_{\mathrm{M}}$ is the LAFE *macroscopic area* (or "footprint"). This parameter $\alpha_{\mathrm{f}}^{\mathrm{SN}}$ is a measure of what proportion of the emitter surface is actually emitting electrons. It is not an accurate measure, but looks a useful parameter when making comparisons in emitter research and development.

Formal area efficiencies have been calculated from measurements on several typical FE film cathodes. Results summarized in Table 3 suggest that emitter performance could probably be improved if fabrication procedures could be optimized.

| Table 3: Formal emission area and formal area efficiency (derived by assuming a SN barrier), for several film-type large-area field electron emitters. | | | |
|---|---|---|---|
| Material and reference | CNT [47] | ZnO [21] | Mo [48] |
| Work function (eV) | 4.6 | 3.3 | 4.7 |
| Substrate area $A_{\mathrm{M}}$ (mm$^2$) | 0.92 | 500 | 6.25 |
| Extracted formal emission area $A_{\mathrm{f}}^{\mathrm{SN}}$ (nm$^2$) | $8\times10^6$ | $2.3\times10^4$ | $5.5\times10^3$ |
| Extracted formal area efficiency $\alpha_{\mathrm{f}}^{\mathrm{SN}}$ | $8.70\times10^{-6}$ | $4.6\times10^{-11}$ | $8.8\times10^{-10}$ |

4.2 Availability of computer program

To get good results for the extracted formal emission area $A_{\mathrm{f}}^{\mathrm{SN}}$, an iteration process is needed. For an assumed value of fitting voltage $V_{\mathrm{t}}$, this iteration finds a self-consistent value for the voltage-to-local-field conversion factor $\beta$. The iteration loop is illustrated in Fig. 5. The parameter $\mathrm{s_t}$ is the fitting value of the slope correction factor, "$i$" is the iteration label, and $\mathcal{F}$ and $\mathcal{G}$ are functions defined in the ESM.

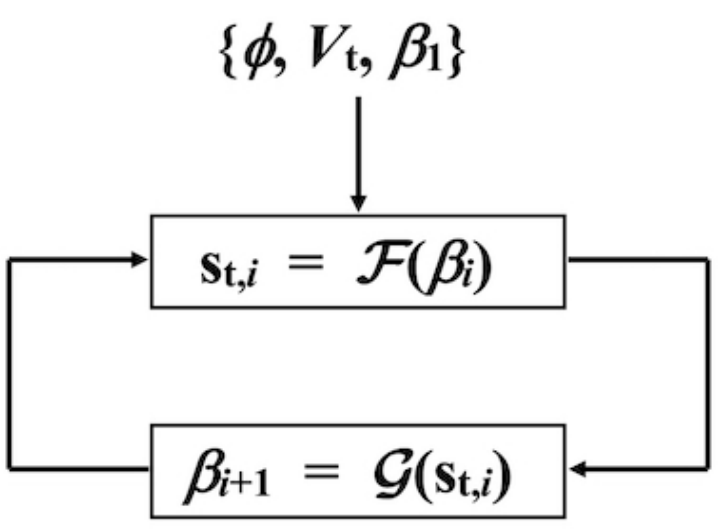

Fig. 5. To show structure of the iteration loop used for finding precise estimates of the parameters $f_{\mathrm{t}}$ and $\mathrm{s_t}$, and of the conversion factor $\beta$, for a given value of fitting voltage $V_{\mathrm{t}}$.

A program has been built that takes the necessary input data including the parameters $S^{\mathrm{expt}}$ and $R^{\mathrm{expt}}$, and carries out this iteration. For selected values of $V_{\mathrm{t}}$ it generates corresponding estimates of $f_{\mathrm{t}}$, $\mathrm{s_t}$, $\beta$, $F_{\mathrm{t}}$, $\mathrm{r_t}$, $A_{\mathrm{f}}^{\mathrm{SN}}$, $d_{\mathrm{F}}$ and $B_{\mathrm{pract}}$. A second part of the program employs user-inputs of local work function, temperature and field, and equations given by Swanson and Schwind [10], to calculate the related total-energy distribution, as shown in Fig. 4(a).

More information about the program is provided in the ESM. This program is freely available. The related .exe file and "users' manual" can be downloaded from https://www.tangshuai-fe-sysu.org.cn/survey.html. We hope that the availability of this easy-to-use program will encourage experimental FE researchers to use EMG FE theory, rather than elementary FE theory, to analyse FE experimental current-voltage data plotted as a FN plot.

### 4.3 Strength of FIM-FEM method and limitations of MG FE theory

Finally, we note that EMG FE theory is a planar FE theory that also disregards the role of atomic-level wave-functions. It will eventually be necessary to go beyond MG FE theory, as partially demonstrated by the work [49, 50] of Kyritsakis and colleagues on FE from Earthed spherical emitters. If practical emitters become ultrasharp or atomic-scale, then existing plot-based data-analysis methods will not work. However, the FIM-FEM method can be used to obtain the electron source area even at the near-atomic scale. Thus, this method is expected to be useful for characterizing upcoming atomic-scale sharply pointed electron sources.

## 5. Conclusions

This paper has introduced a new FIM-FEM experimental method for finding the electron source area $A^{\mathrm{S}}$ of an operating field electron emitter. It finds that $A^{\mathrm{S}}$ is much closer to the formal emission area $A_{\mathrm{f}}^{\mathrm{SN}}$ derived in FN-plot analysis by assuming a SN tunnelling barrier than to the formal emission area $A_{\mathrm{f}}^{\mathrm{ET}}$ derived by assuming an ET barrier. This seems welcome additional experimental confirmation that MG FE theory (based on the SN barrier) is better physics than elementary FE theory or 1920s FE theory, both based on the ET barrier, and that $A_{\mathrm{f}}^{\mathrm{SN}}$ is the better estimate of emission area.

The "messy" procedure needed when using MG FE theory to analyse FN plots has been packaged into a simple program that implements an iterative method. For various assumed fitting values of measured voltage, this program can extract/calculate important characteristics of FE sources, including apex field strength, apex FEF, formal emission area, the total-energy distribution (and hence its full width at half-maximum), and source practical brightness.

The program also determines the formal area efficiency of a large-area field electron emitter: this provides a measure of the efficiency of a film cathode. We hope that the availability of an easy-to-use program will encourage experimental FE researchers to use EMG FE theory, rather than elementary FE theory, to analyse FE experimental current-voltage data plotted as a FN plot. Finally, as already noted, because our new FIM-FEM method will work for apex-radius values smaller than those at which FN-plot-based methods become invalid, we expect our new method to be useful for characterizing upcoming atomic-scale sharply pointed electron sources.

## 6. Acknowledgement

This work was supported by the National Key Basic Research Program of China (Grant no. 2024YFA1209004), the National Natural Science Foundation of China (Grant no. 62301619, 62274188), the Key field Research Program of Guangdong Province (Grant no.

2022B0303030001), the Special Topic on Basic and Applied Basic Research of Guangzhou (Grant no. Grant no. 2024A04J4525), the Guangdong Provincial Association for Science and Technology Young Scientific and Technological Talent Cultivation Program, the Fundamental Research Funds for the Central Universities, Sun Yat-sen University (Grant no. 24hytd002-2) and start-up funds for the Central Universities, Sun Yat-sen University.

## Supplementary Materials

We provide as Electronic Supplementary Material a detailed account of the present state of that part of field electron emission theory that is directly relevant to the main paper, as well as the evaluation of the fitting parameters of computer program.

## Contributions

S.Z. Deng, N.S. Xu and S. Tang conceived and designed the study. S. Tang designed and performed the FIM-FEM experiments and electron-optic simulations. S. Tang and M.K. Gou write the program. M.K. Gou and Y.Z. Hu helped with electron-optic simulations. J. Tang and L.-C. Qin helped with FIM-FEM experiments. Y. Zhang and Y. Shen helped with results discussion. R.G. Forbes provided theoretical guidance about MG and FN theory. S. Tang, M.K. Gou, Y.Z. Hu, N.S. Xu, R.G. Forbes, S.Z. Deng analysed the data. S. Tang write the original paper and R.G. Forbes and S.Z. Deng revised the paper with input from all authors.

## Declaration of competing interest

The authors declare that they have no known competing financial interests or personal relationships that could have appeared to influence the work reported in this paper.

## Data availability

The data that support the findings of this study are available from the corresponding authors upon reasonable request.

# Methods for characterization of atomic-scale field emission point-electron-source

Shuai Tang[1], Mingkai Gou[1], Yingzhou Hu[1], Jie Tang[2], Yan Shen[1], Yu Zhang[1], Lu-chang Qin[2], Ningsheng Xu[1], Richard G. Forbes[3], Shaozhi Deng[1]*

[1]State Key Laboratory of Optoelectronic Materials and Technologies, Guangdong Province Key Laboratory of Display Material and Technology, School of Electronics and Information Technology, Sun Yat-sen University, Guangzhou 510275, China

[2]Faculty of Materials and Manufacturing, Beijing University of Technology, Beijing 100124, China

[3]School of Mathematics and Physics, University of Surrey, Guildford, Surrey GU2 7XH, United Kingdom

Corresponding Author E-mail: stsdsz@mail.sysu.edu.cn

**ELECTRONIC SUPPLEMENTARY MATERIAL**

**Background theoretical information**

## 1. Introduction

The aim of this supplementary document is to bring together in a single place most of the background material needed to fully understand the calculations in this paper, with appropriate references to original derivations and discussions. The material here includes: theoretical data and equations; discussions of how the various parameters involved are derived and defined; discussion of how characterisation parameters are extracted; and some other relevant material.

In order to make a distinction between the electric fields associated with Coulomb-type electrostatics and the electrical component of a travelling electromagnetic wave, a Coulomb-type electrical field is always called here an *electrostatic (ES) field.*

As in the main paper, this document uses the so-called *electron emission sign convention*, in which electrostatic fields, electron currents and electron current densities are regarded as positive, even though they would be negative in conventional textbook electrostatics. The symbol $F$ is used to denote the negative of conventional textbook electrostatic field: $F$ is thus positive in most circumstances affecting field electron emission (FE). Strictly, the conventional classical electrostatic name for $F$ is "electrostatic potential gradient (ESPG)", but this note follows the usual FE-literature convention of calling $F$ the "field".

As in the main paper, all equations are defined within the modern system, introduced in the 1970s and now called the *International System of Quantities (ISQ)*, in which the *electric constant* $\varepsilon_0$ appears in Coulomb's Law. However, extensive use is made of so-called *FE customary units*. These customary units are fully and permanently compatible with ISQ equations. The main features of customary units are that: energy is expressed in eV; charge is expressed in eV $V^{-1}$; current is expressed in A; distance is expressed in nm or m (depending on the context); and mass is expressed in eV $nm^{-2}$ $s^2$.

In these notes, electron emission induced by a negative ES field is termed "field electron emission". Other names in literature use are "field emission" and "electron field emission". The abbreviation "FE" is regarded as standing for any of these equivalent names.

## 2. Fundamental physical constants

FE theory makes use of the following fundamental constants: the elementary (positive) charge, $e$; the electric constant (aka the vacuum electric permittivity), $\varepsilon_0$; the electron rest mass in free space, $m_e$; Planck's constant, $h_P$; Planck's constant divided by $2\pi$, $\hbar$; and the Boltzmann constant, $k_B$. The May 2019 values [R1] of these constants have been used. The Electronic Supplementary Material (ESM2) to Ref. [R2] records values of these constants, both in SI units and in FE customary units.

## 3. Universal constants and related parameters

The following universal constants and related parameters, tabulated in [R2], form part of the background to the main paper.

The Schrödinger-equation constant for an electron ($\kappa_e$), the Sommerfeld electron supply constant ($z_S$), first ($a$) and second ($b$) Fowler-Nordheim (FN) constants, the factor $4\pi\varepsilon_0$ and the Schottky constant ($c$) are universal constants defined and given (to 7 sig. figs) as follows:

$$\kappa_e \equiv (2m_e)^{1/2}/\hbar \cong 5.123168 \text{ eV}^{-1/2}\text{ nm}^{-1}; \quad \text{(E1)}$$

$$z_S \equiv 4\pi e m_e/h_P^3 \cong 1.618311\times10^{14}\text{ A m}^{-2}\text{ eV}^{-2}; \quad \text{(E2)}$$

$$a \equiv e^3/8\pi h_P \cong 1.541453\ \mu\text{A eV V}^{-2}; \quad \text{(E3)}$$

$$b \equiv (4/3)(2m_e)^{1/2}/e\hbar \cong 6.830890 \text{ eV}^{-3/2}\text{ V nm}^{-1}; \quad \text{(E4)}$$

$$4\pi\varepsilon_0 \cong 0.694\,4615 \text{ eV V}^{-2}\text{ nm}; \quad \text{(E5)}$$

$$c \equiv (e^3/4\pi\varepsilon_0)^{1/2} \cong 1.199985 \text{ eV V}^{-1/2}\text{ nm}^{1/2}. \quad \text{(E6)}$$

The constant $\kappa_e$ was defined by Fowler and Nordheim [R3]. The constant $z_S$ is associated with the statistical mechanics of the Sommerfeld model (e.g., see [R4]). The constant $b$ emerges when determining the transmission probability for a triangular or approximately triangular potential-energy (PE) barrier (see Appendix A1). The constant $a$ then emerges when a double integration is made over the energies of free-electron-metal internal energy states, in order to derive an expression for local emission current density (LECD) (see Section 11.1 below)). The factor $4\pi\varepsilon_0$ is internationally defined to be the constant that appears in the ISQ version of Coulomb's law. The constant $c$ arises in connection with the Schottky-Nordheim barrier (see below) via the equation

$$\Delta_S = cF^{1/2}, \quad \text{(E7)}$$

where $\Delta_S$ is the "Schottky reduction" (of the barrier to electron escape) caused by an ES field of magnitude $F$.

The following combinations of universal constants are useful:

$$2\kappa_e \cong 10.24624 \text{ eV}^{-1/2}\text{ nm}^{-1}; \quad \text{(E8)}$$

$$ab^2 \cong 7.192493 \text{ A eV}^{-2}\text{ nm}^{-2}; \quad \text{(E9)}$$

$$2/3b \cong 9.759588\times10^{-2}\text{ eV}^{3/2}\text{ (V/nm)}^{-1}; \quad \text{(E10)}$$

$$c^2 \cong 1.439965 \text{ eV}^2\text{ V}^{-1}\text{ nm}; \quad \text{(E11)}$$

$$bc^2 \cong 9.836239 \text{ eV}^{1/2}. \quad \text{(E12)}$$

The following parameters, involving the local work function $\phi$, need to be evaluated:

$$\text{Scaling parameter, } \eta \equiv bc^2\phi^{-1/2} \cong 9.836\cdot(\text{eV}/\phi)^{1/2}. \quad \text{(E13)}$$

$$\text{Combination, } ab^2\phi^2 \cong 7.1925 \cdot (\phi/\text{eV})^2 \cdot (\text{A nm}^{-2}) \,. \qquad \text{(E14)}$$

The accuracy of these last two parameters depends on the accuracy with which $\phi$ is known.

## 4. Smooth planar metal emitter (SPME) methodology

The simplest theoretical treatments of FE use planar models that aim to describe FE from metals. Emission is modelled as taking place from the smooth planar surface of a Sommerfeld-type metal model. Atomic structure and atomic-level wave-functions are disregarded. This approach is sometimes called *smooth planar metal emitter (SPME) methodology*.

## 5. Basic definitions relating to electron energies

Denote the direction normally outwards from the Sommerfeld model surface by $z$, and call the electron-energy component associated with motion in the $z$-direction the *electron normal-energy*. The electron-energy component parallel to the model surface is purely kinetic in nature and is termed the *electron parallel-energy*. The *electron total-energy* is the sum of these components.

In FE theory it is convenient to measure both total-energy and normal-energy from the emitter Fermi level. Energies measured in this way are denoted by $\varepsilon_{\text{tot}}$ and $\varepsilon_{\text{n}}$, respectively. Parallel energy is denoted by $\varepsilon_{\text{p}}$. Clearly, $\varepsilon_{\text{tot}}=\varepsilon_{\text{n}}+\varepsilon_{\text{p}}$. For consistency, *electron potential energy* (PE), denoted here by $U_{\text{e}}$, must also be measured from the Fermi level. Figure S1 shows electron normal-energy plotted against distance $z$, and illustrates some of these definitions. It also shows the local work function $\phi$ and the *zero-field barrier height* $H$ applicable to an electron approaching the model surface with normal energy $\varepsilon_{\text{n}}$. Clearly $H=\phi-\varepsilon_{\text{n}}$.

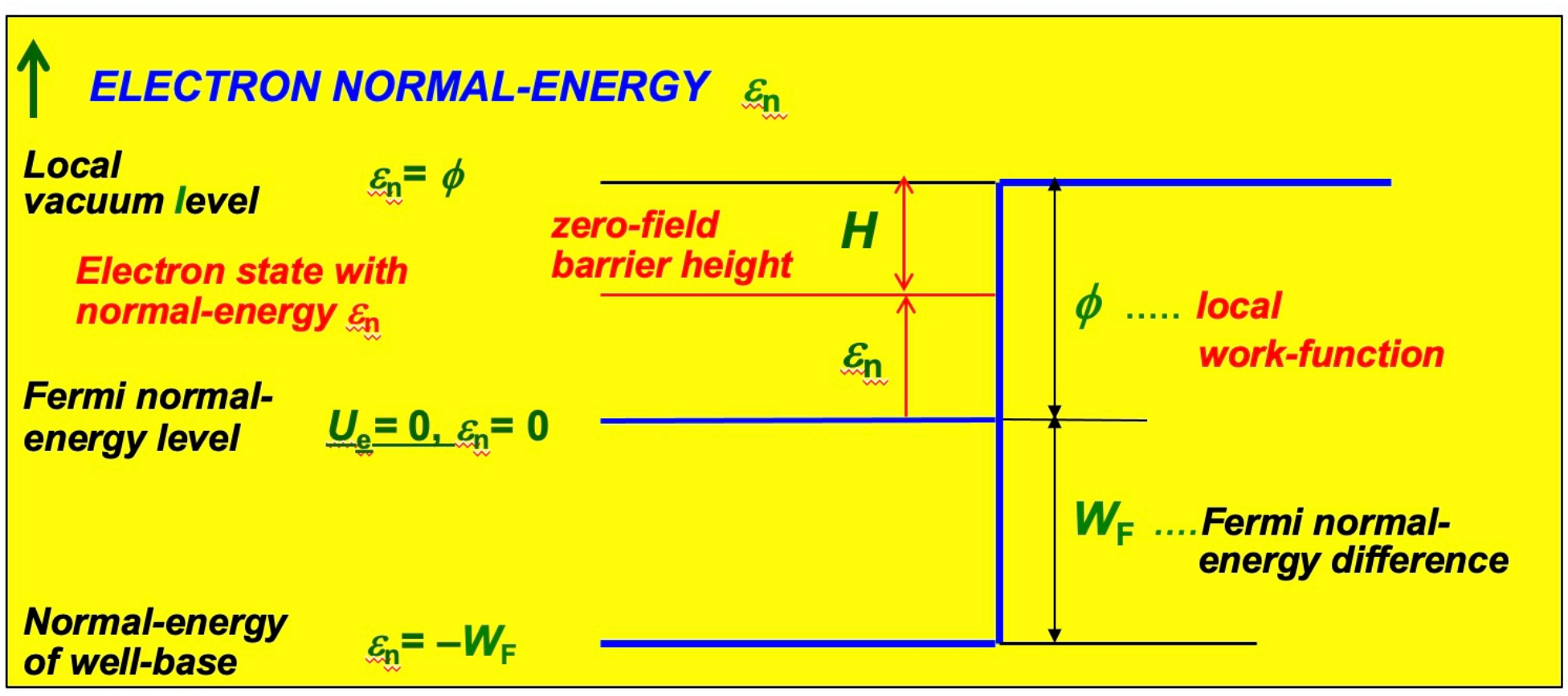

Fig. S1: To illustrate definitions of energies. Note that: $H\equiv\phi-\varepsilon_{\text{n}}$.

## 6. Motive energies for the "exactly triangular" (ET) and "Schottky-Nordheim" (SN) barriers

In this planar geometry, electron transmission barriers are defined by the difference between the electron potential energy $U_{\text{e}}(z)$ and the electron normal-energy $\varepsilon_{\text{n}}$. This energy difference is called here the *electron motive energy* and denoted by $M(H,F,z)$.

For local surface "field" $F$, the motive energy $M^{\text{ET}}(H,F,z)$ for an *exactly triangular (ET) PE barrier of zero-field height $H$* is given by

$$M^{\text{ET}}(H,F,z) \equiv H - eFz \,. \qquad \text{(E15)}$$

The ET barrier takes electrostatic effects into account, but disregards an exchange-and-correlation (E&C) interaction between the escaping electron and the emitter. When this E&C interaction is modelled by a classical image PE, the motive energy $M^{SN}(H,F,z)$ for the resulting *Schottky-Nordheim (SN) PE barrier* is given by

$$M^{SN}(H,F,z) \equiv H - eFz - e^2/16\pi\varepsilon_0 z\ . \qquad \text{(E16)}$$

Figure 2 below illustrates these barriers. For the special case where the zero-field barrier height is equal to the local work function $\phi$, the parameter $H$ in the above equations is replaced by $\phi$.

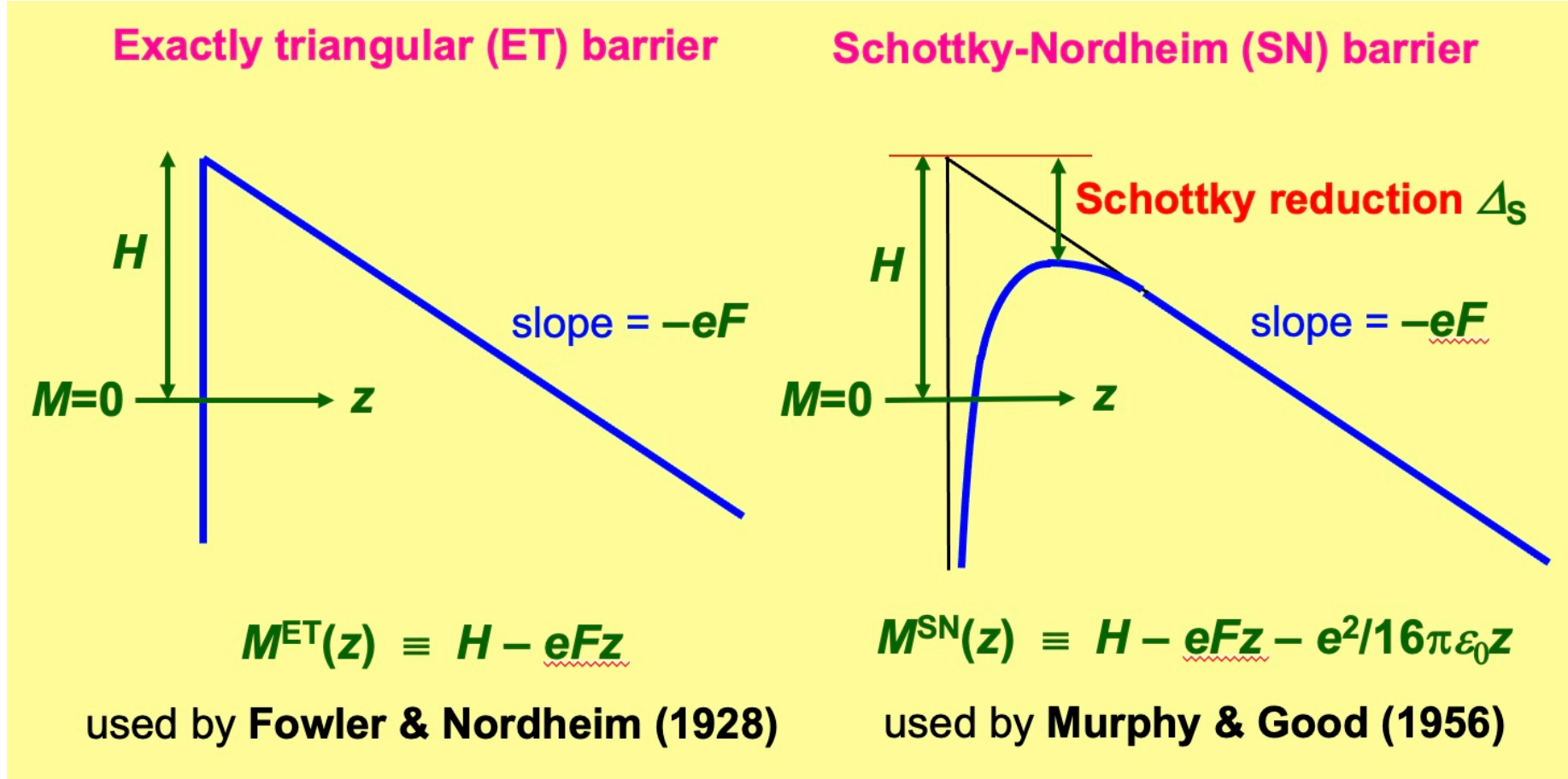

Fig. S2: To illustrate the forms of the ET and SN PE barriers.

## 7. Schottky reduction, reference field, scaled field and Nordheim parameter

As noted above, the effect of the image-PE term in eq. (E16) is to reduce the height of the zero-field barrier by an amount $\Delta_S$ called here the *Schottky reduction* and given by

$$\Delta_S = cF^{1/2}, \qquad \text{(E17)}$$

where $c$ is the *Schottky constant* defined above.

In the case of a SN barrier of zero-field height $H$, the local "field" $F_H$ necessary to reduce the barrier to zero is obtained by setting $\Delta_S = H$, yielding

$$F_H = c^{-2}H^2. \qquad \text{(E18)}$$

The so-called *local scaled field* ($f_H$) *for a SN barrier of zero-field height H* is a positive dimensionless parameter that is related to the local "field" $F$ by

$$f_H \equiv F/F_H = c^2 H^{-2} F\ . \qquad \text{(E19)}$$

In the case of a SN barrier of zero-field height $\phi$, the "field" $F_\phi$ is the "field" necessary to pull the top of the SN barrier down to the Fermi level. This field has also been denoted by the symbol $F_R$ and called the *reference field*. The related *scaled field for a SN barrier of zero-field height* $\phi$ is a positive dimensionless parameter given by

$$f_\phi \equiv F/F_\phi = c^2 \phi^{-2} F\ . \qquad \text{(E20)}$$

When no confusion is possible, the subscript "$_\phi$" is normally dropped. In what follows, plain "$f$" always means "$f_\phi$". This scaled field for a SN barrier of zero-field height $\phi$ plays an important role in modern FE theory.

Instead of scaled fields, many treatments of the SN barrier in FE literature (particularly older treatments) use the *Nordheim parameters* $y$ and $y_F$, *for SN barriers of zero-field height H and* $\phi$, respectively. These parameters are defined as follows:

$$y \equiv cH^{-1}F^{1/2} = f_H^{1/2}; \qquad \text{(E21)}$$

$$y_F \equiv c\phi^{-1}F^{1/2} = f_\phi^{1/2}. \qquad \text{(E22)}$$

As indicated below, use of scaled fields rather than Nordheim parameters is considered mathematically superior. Scaled fields are used in the modern theory presented here.

## 8. Gamow factor and the simple-JWKB formalism for transmission probability

***Terminology.*** The probability that an electron approaching the emitter surface from the inside will escape, either by tunnelling through a surface barrier or by passage over the top of the barrier, is called here the *transmission probability*. This term is now in strongly preferred over the term "transmission coefficient" occasionally used, because this latter name is ambiguous, in that it is also sometimes used to refer to the transmission amplitude. The general notation "$D$" for transmission probability, which was used by Fowler and Nordheim [R3] and by Landau and Lifschitz [R5] is preferred over the notation "$T$" sometimes used (particularly in theoretical papers), since this allows the symbol $T$ to be used for thermodynamic temperature.

***Gamow factor.*** For a PE barrier defined by $H$ and $F$, the modern *Gamow factor* $G(H,F)$ [R6,R7] is a dimensionless mathematical-modelling parameter defined in terms of $M(H,F,z)$ by

$$G(H,F) \equiv 2\kappa_e \int \{M(H,F,z)\}^{1/2} dz \,, \tag{E23}$$

where the integral is taken "across the barrier", in the range of $z$ where $M \geq 0$, i.e. between the zeros of $M(H,F,z)$.

In older work, an universal constant $g_e$ was defined by $g_e \equiv 2\kappa_e$ and appeared in eq. (E23), but the introduction of an extra symbol was subsequently considered unnecessary.

Obviously, the precise mathematical form of $G(H,F)$ will depend on the barrier form chosen and hence on the precise mathematical form of the motive energy $M(H,F,z)$. However, some mathematical results apply to all planar-barrier forms and it is convenient to first state results for a *"general planar barrier"* (GPB).

Given an expression for $G(H,F)$, there are four different mathematical formalisms ("*transmission formalisms*") [R8] that could be used to derive an expression for the barrier transmission probability $D(H,F)$. The simplest of these is the so-called *simple-JWKB (sJWKB) formalism* (also called the "first-order JWKB formalism"). In this formalism the transmission probability for a barrier of zero-field height $H$ is given by

$$D(H,F) \sim \exp[-G(H,F)] \,. \tag{E24}$$

All widely used versions of planar FE theory use this sJWKB formalism in their derivations.

In FE, $G(H,F)$ has also been called the "barrier strength" [high $G$ = strong barrier = low $D$].

## 9. Decay width at the (zero-temperature) Fermi level

An important parameter in FE theory, in particular in relation to the optical properties of needle-like FE sources, is the so-called *decay-width at the (zero-temperature) Fermi level*. This parameter is denoted here by $d_F$, though often just by "$d$" in the literature (and sometimes in other ways). Although algebraic expressions for $d_F$ (or $d_F^{-1}$) can be found in the literature, it is difficult to find a precise definition of what the symbol $d_F$ is intended to represent.

The authors' view is that (in planar FE theory) $d_F$ should in principle be defined by the equation:

$$d_F^{-1} \equiv (\partial \ln D/\partial \varepsilon_n)_F|_{(\varepsilon_n=0)} = -(\partial \ln D/\partial H)_F|_{(H=\phi)}. \tag{E25}$$

When defined in this way, the theoretically estimated value of $d_F$ depends both on the form assumed for the tunnelling barrier and on the formalism used to estimate transmission probability. Thus, when the Landau & Lifshitz (LL) formalism [R8] is applied to the general planar barrier (GPB), the transmission probability $D(H,F)$ and the corresponding estimate of $d_F$ are given (at a general level) by

$$D(H,F)[\text{LL+GPB}] = P^{\text{GPB}} \exp[-G^{\text{GPB}}] \,, \tag{E26}$$

$$d_F^{-1}[\text{LL+GPB}] = -(\partial \ln P^{\text{GPB}}/\partial H)_F|_{(H=\phi)} + (\partial G^{\text{GPB}}/\partial H)_F|_{(H=\phi)} \,, \tag{E27}$$

where $P^{\text{GPB}}$ is a pre-factor of order unity. For the ET barrier, an expression for $P^{\text{ET}}$ has been derived (e.g., [R3,R9]). For the SN barrier, $P^{\text{SN}}$ can be estimated numerically [R10], but (at present) there is no known analytical form. The situation as regards other barriers is not examined here.

However, if the usual sJWKB formalism is used instead, then the prefactor term is not present and we get

$$D(H,F)[\text{sJWKB+GPB}] = \exp[-G^{\text{GPB}}(H,F)] \,, \tag{E28}$$

$$D(\phi,F)[\text{sJWKB+GPB}] = \exp[-G^{\text{GPB}}(\phi,F)] \,, \tag{E29}$$

$$d_F^{-1}[\text{sJWKB+GPB}] = (\partial G^{\text{GPB}}/\partial H)_F|_{(H=\phi)} \,. \tag{E30}$$

In FE literature, this last expression is what is normally used to define "decay width", and is what will be used here. In what immediately follows, the simpler notation "$d_F^{\text{GPB}}$" will be used.

It needs to be remembered, however, that decay-width values predicted in this way are approximations. These approximate values are expected to be "more than good enough for most current technological purposes", but there is no good knowledge of precisely how big any errors might be.

## 10. General-form equation for local emission current density in "deep-tunnelling" conditions

The term *"in deep-tunnelling conditions"* implies that most or nearly all electrons escape by tunnelling at normal-energy levels significantly below the top of the PE barrier.

Derivations of related equations use: (a) the electron statistical mechanics of the Sommerfeld model (including Fermi-Dirac statistics); (b) either the Landau&Lifshitz or the simple-JWKB transmission formalism (but usually the latter); and (c) a specific barrier form. The s-JWKB formalism will be used here. However, statistical aspects of the theory apply to all barrier forms, so initially theory will be developed for a general planar barrier (GPB).

### *10.1 Electron statistical behaviour in the Sommerfeld model*

Consider a geometrical plane inside a metal modelled by the Sommerfeld model, and consider electrons approaching the plane that have energy components $\varepsilon_n$ and $\varepsilon_p$ normal and parallel to the plane, respectively. It can be shown (e.g., [R11]) that the contribution $d^2Z_{\text{sup}}$, to the electron-current density (considered positive) approaching the plane, made by electrons in the small energy ranges $d\varepsilon_n$, $d\varepsilon_p$, is

$$d^2Z_{\text{sup}} = f_{\text{F-D}}(T,\varepsilon_{\text{tot}}) \cdot z_S \, d\varepsilon_n, d\varepsilon_p \,, \tag{E31}$$

where $z_S$ is the universal constant, defined above, now called the *Sommerfeld electron supply constant*, and $f_{\text{F-D}}(T,\varepsilon_{\text{tot}})$ is the *Fermi-Dirac distribution function,* where $T$ is *thermodynamic temperature*. In the present context, $f_{\text{FD}}(T,\varepsilon_{\text{tot}})$ is conveniently expanded in the form

$$f_{\text{F-D}}(T,\varepsilon_{\text{tot}}) = 1/[1+\exp\{(\varepsilon_n+\varepsilon_p)/k_BT\}] \,. \tag{E32}$$

Each electron state makes a contribution to the *local emission current density (LECD)* $J_L(\phi,T,F)$. Thus, a two-dimensional integration of $d^2Z_{\text{sup}}$ over $\varepsilon_n$ and $\varepsilon_p$ is needed. However, since (in a free-electron model) the transmission probability depends on normal-energy but not on parallel-energy, the double integral over energy can be split into two steps, as follows.

### *10.2 Integration over parallel-energy to give $N(T,\varepsilon_n)$*

First, an integral over $\varepsilon_p$ is performed to give a function denoted below by $N(T,\varepsilon_n)$. (Surprisingly, the integral does have an exact analytical solution, as may be shown by differentiating the expression in the second equation below.)

$$N(T,\varepsilon_n) = z_S \int_0^\infty 1/[1+\exp\{(\varepsilon_n+\varepsilon_p)/k_BT\}] d\varepsilon_p$$

$$= \; -z_{\mathrm{S}}k_{\mathrm{B}}T\left[\ln\left[1+\exp\left\{(-\varepsilon_{\mathrm{n}}-\varepsilon_{\mathrm{p}})/k_{\mathrm{B}}T\right\}\right]\right]_{\varepsilon_{\mathrm{p}}=0}^{\varepsilon_{\mathrm{p}}=\infty}$$

$$= \; z_{\mathrm{S}}k_{\mathrm{B}}T\ln[1+\exp\{-\varepsilon_{\mathrm{n}}/k_{\mathrm{B}}T\}]. \tag{E33}$$

This function $N(T,\varepsilon_{\mathrm{n}})$ is the incident electron-current density per unit normal-energy range, and is called here the *incident normal-energy distribution (i-NED)*.

Older discussions of emission theory used a slightly different function, denoted here by $N$(old), that is related to $N(T,\varepsilon_{\mathrm{n}})$ by

$$N(\mathrm{old}) = N(T,\varepsilon_{\mathrm{n}})/e\,. \tag{E34}$$

This function $N$(old) was called the "supply function". It is considered better modern practice (more consistent with other aspects of emission theory) to "put $e$ into $N$", rather than show it separately.

### *10.4 Integration over normal-energy to give* $J_{\mathrm{L}}(\phi,T,F)$

The following integral over normal-energy then gives the LECD for a general planar barrier:

$$J_{\mathrm{L}}^{\mathrm{GPB}}(\phi,T,F) = \int_{-W_{\mathrm{F}}}^{\infty} N(T,\varepsilon_{\mathrm{n}})\cdot D^{\mathrm{GPB}}(\phi,F,\varepsilon_{\mathrm{n}})\,\mathrm{d}\varepsilon_{\mathrm{n}}\,, \tag{E35}$$

where, as shown in Fig. S1, $W_{\mathrm{F}}$ is a positive parameter that is the difference in normal-energy between the Fermi level and the base of the Sommerfeld well. In most practical cases, the transmission-probability expression is very small for $\varepsilon_{\mathrm{n}}$-vales near $(-W_{\mathrm{F}})$ and below, so with negligible error the lower limit of integration can be taken as $-\infty$.

To carry out the integration and obtain an algebraic result, usual practice is to formally Taylor-expand $G^{\mathrm{GPB}}$ about the Fermi level, and retain only the first-order term. This yields

$$D^{\mathrm{GPB}}(\phi,F,\varepsilon_{\mathrm{n}}) \;\approx\; \exp[-G^{\mathrm{GPB}}(\phi,F,0) - (\partial G^{\mathrm{GPB}}/\partial\varepsilon_{\mathrm{n}})_{\phi,F}\cdot\varepsilon_{\mathrm{n}}]\,, \tag{E36}$$

$$D^{\mathrm{GPB}}(\phi,F,\varepsilon_{\mathrm{n}}) \;\approx\; D_{\mathrm{F}}^{\mathrm{GPB}}(\phi,F)\exp[\varepsilon_{\mathrm{n}}/d_{\mathrm{F}}^{\mathrm{GPB}}]\,, \tag{E37}$$

where subscript "F" labels a value "taken at the Fermi level" (or, equivalently, taken for a barrier of zero-field height $\phi$), and parametric dependences are omitted for notational simplicity.

With all the above approximations, eq. (E35) becomes

$$J_{\mathrm{L}}^{\mathrm{GPB}}(\phi,T,F) = z_{\mathrm{S}}D_{\mathrm{F}}^{\mathrm{GPB}}\int_{-\infty}^{\infty} k_{\mathrm{B}}T\ln[1+\exp\{-\varepsilon_{\mathrm{n}}/k_{\mathrm{B}}T\}]\cdot\exp[\varepsilon_{\mathrm{n}}/d_{\mathrm{F}}^{\mathrm{GPB}}]\,\mathrm{d}\varepsilon_{\mathrm{n}}. \tag{E38}$$

This integral can be solved exactly by first integrating by parts (which reduces it to a standard form), and then applying a formula from a handbook of standard integrals. The details are lengthy and slightly messy, cannot easily be found in published literature, and are recorded elsewhere [R12]. The outcome is a simple "general" (or "abstract") planar formula for the LECD, namely

$$J_{\mathrm{L}}^{\mathrm{GPB}}(\phi,T,F) = \lambda_{T}^{\mathrm{GPB}}z_{\mathrm{S}}\cdot(d_{\mathrm{F}}^{\mathrm{GPB}})^{2}D_{\mathrm{F}}^{\mathrm{GPB}}, \tag{E39}$$

where the *temperature correction factor* $\lambda_{T}^{\mathrm{GPB}}$ *for the general planar barrier* is given by

$$\lambda_{T}^{\mathrm{GPB}} \;=\; \pi p^{\mathrm{GPB}}/\sin(\pi p^{\mathrm{GPB}}), \tag{E40}$$

and the *Swanson-Bell parameter* $p^{\mathrm{GPB}}$ *for the general planar barrier* is given by

$$p^{\mathrm{GPB}} \equiv k_{\mathrm{B}}T/d_{\mathrm{F}}^{\mathrm{GPB}}. \tag{E41}$$

In the limit of zero temperature $\lambda_{T}^{\mathrm{GPB}} \to 1$, and the zero-temperature form of the GPB equation for LECD is obtained.

Note that the form of (E39) allows a simple physical explanation of this equation. The first three terms can be collected into an *effective local incident electron-current density*, $Z_{\mathrm{F}}^{\mathrm{GPB}}$, thus

$$Z_{\mathrm{F}}^{\mathrm{GPB}} \;=\; \lambda_{T}^{\mathrm{GPB}}z_{\mathrm{S}}\cdot(d_{\mathrm{F}}^{\mathrm{GPB}})^{2}\,, \tag{E42}$$

enabling eq. (E39) to be written in the "abstract" form

$$J_{\mathrm{L}}^{\mathrm{GPB}}(\phi,T,F) = Z_{\mathrm{F}}^{\mathrm{GPB}} D_{\mathrm{F}}^{\mathrm{GPB}}. \tag{E43}$$

Equation (E43) shows that the LECD is given by the product of the "transmission probability at the Fermi level" and the "effective local incident electron-current density"; eq. (E42) shows that $Z_{\mathrm{F}}^{\mathrm{GPB}}$ is the product of the "incident electron-current density per unit area of ($\varepsilon_{\mathrm{n}}$,$\varepsilon_{\mathrm{n}}$) energy space $z_{\mathrm{S}}$" and the "effective area of energy space $\lambda_{T}^{\mathrm{GPB}} \cdot (d_{\mathrm{F}}^{\mathrm{GPB}})^2$" from which the emitted electron-current is drawn.

Also note that using only the first-order term in Taylor expansion (E36) is an approximation that gets increasingly weak as $|\varepsilon_{\mathrm{n}}|$ increases. The issues involved have been discussed by Gadzuk and Plummer [R13]. In particular, see their Fig. 4.

## 11. The LECD equations for the ET and SN barriers

### *11.1 Derivation of the elementary FE equation for LECD*

The elementary (EL) FE equation is based on the s-JWKB formalism and the ET barrier. From equations above, we find that

$$G^{\mathrm{ET}}(H,F) = bH^{3/2}/F\ ; \tag{E44}$$

$$G_{\mathrm{F}}^{\mathrm{ET}} = b\phi^{3/2}/F\ ; \tag{E45}$$

$$D_{\mathrm{F}}^{\mathrm{ET}} = \exp[-b\phi^{3/2}/F]\ ; \tag{E46}$$

$$d_{\mathrm{F}}^{\mathrm{ET}} = (2/3b)\phi^{-1/2}F\ ; \tag{E47}$$

From (E2) and (E47) we find (after some algebraic manipulation):

$$z_{\mathrm{S}}\cdot(d_{\mathrm{F}}^{\mathrm{ET}})^2 \ = \ (4z_{\mathrm{S}}/9b^2)\phi^{-1}F^2 \ = \ (e^3/8\pi h_{\mathrm{P}})\,\phi^{-1}F^2 \ \equiv \ a\phi^{-1}F^2\ . \tag{E48}$$

where $a$ is the first FN constant, as defined above. The elementary (zero-temperature) FE equation is thus given by

$$J_{\mathrm{L}}^{\mathrm{EL}}(\phi,F) = a\phi^{-1}F^2\exp[-b\phi^{3/2}/F]. \tag{E49}$$

There does exist a finite-temperature version of the elementary equation, obtained by multiplying eq. (E49) by an ET version of factor (E40), but nowadays this is rarely or never used.

### *11.2 The "21st Century" version of the 1956 Murphy-Good FE equation for LECD*

Murphy-Good (MG) type equations are based the s-JWKB formalism and the SN barrier. Evaluation of integral (E23) for the SN barrier leads to a result of the form

$$G^{\mathrm{SN}}(H,F) \ = \ \mathrm{v}(H,F)\cdot bH^{3/2}/F\ , \tag{E50}$$

where v($H$,$F$) is a mathematical correction factor.

This factor v($H$,$F$) can be expressed in terms of a single "auxiliary variable". This can be either the variable $y$ defined above, which is the so-called "legacy approach", or the variable $f_H$ defined above, which is the so-called "21st Century approach". The "legacy approach" was developed in the 1950s by Murphy and Good [R14], following earlier work by Burgess, Kroemer and Houston (BKH) [R15] and Dyke and Trolan [R16], with a useful overview then given by Good and Müller [R17]. The "21st Century approach" was developed by Forbes and Deane in a series of papers in 2006 to 2008, including [R18–20], with an overview given later, in [R21].

Although, the "legacy approach" is still in widespread use in FE literature in 2026, the "21st Century approach" is mathematically superior and is used here. Reasons for strongly preferring the modern approach are discussed in [R21]. One objective has been to define and separate the pure mathematics involved, which has several technological applications, from the mathematics specifically involved in modelling electron transmission across the SN barrier.

When the "21st Century approach" is applied to the SN barrier, eq. (E50) is written in the more precise form

$$G^{\mathrm{SN}}(H,F) \;=\; \mathrm{v}_{\mathrm{FD}}(f_H) \cdot bH^{3/2}/F\,, \tag{E51}$$

where $\mathrm{v}_{\mathrm{FD}}(f_H)$ is a particular modelling application of a *special mathematical function* (SMF) $\mathrm{v}_{\mathrm{FD}}(x)$ introduced by Forbes and Deane and discussed further below.

Applying definition (E30) to the SN barrier, noting that $\mathrm{v}_{\mathrm{FD}}(f_H)$ is an indirect function of $H$, yields

$$(d_{\mathrm{F}}^{\mathrm{SN}})^{-1} \;=\; (\partial G^{\mathrm{SN}}/\partial H)_F|_{H=\phi} \;=\{\mathrm{v}_{\mathrm{HD}}(f_H) + (2/3)H\,(\partial \mathrm{v}_{\mathrm{HD}}/\partial H)_F|_{H=\phi}\}\cdot(3/2)bH^{1/2}F^{-1}. \tag{E52}$$

The term $H\,(\partial \mathrm{v}_{\mathrm{HD}}/\partial H)_F$ can be transformed by using the relation

$$H\,(\partial \mathrm{v}_{\mathrm{HD}}/\partial H)_F \;=\; (\partial \mathrm{v}_{\mathrm{HD}}/\partial f_H)_F \;\cdot H\,(\partial f_H/\partial H)_F\,. \tag{E53}$$

From definition (E19) above, we get

$$H\,(\partial f_H/\partial H)_F = -2c^2H^{-2} \;=\; -2\,f_H. \tag{E54}$$

Hence the bracketed term in eq. (52) becomes

$$\{\mathrm{v}_{\mathrm{HD}}(f_H) - (4/3)f_H\,(\partial \mathrm{v}_{\mathrm{HD}}/\partial f_H)_F|_{H=\phi}\}. \tag{E55}$$

By introducing a new SMF $\mathrm{t}_{\mathrm{FD}}(x)$ defined in terms of $\mathrm{v}_{\mathrm{FD}}(x)$ by

$$\mathrm{t}_{\mathrm{FD}}(x) \;\equiv\; \mathrm{v}_{\mathrm{FD}}(x) - (4/3)\,x\,(\partial \mathrm{v}_{\mathrm{HD}}/\partial x)\,, \tag{E56}$$

we see that term (E55) becomes

$$\mathrm{t}_{\mathrm{FD}}(f_H)|_{H=\phi} \;=\; \mathrm{t}_{\mathrm{FD}}(f_\phi)\,. \tag{E57}$$

At this point the subscript on $f_\phi$ can be dropped, using the convention that $f$ means $f_\phi$. Hence $d_{\mathrm{F}}^{\mathrm{SN}}$ is given by

$$d_{\mathrm{F}}^{\mathrm{SN}} = \{\mathrm{t}_{\mathrm{FD}}(f)\}^{-1}\cdot(2/3b)\phi^{-1/2}F \;\;=\;\; \{\mathrm{t}_{\mathrm{FD}}(f)\}^{-1}\cdot d_{\mathrm{F}}^{\mathrm{ET}}\,. \tag{E58}$$

It is then clear, from comparisons with equations above, that the "21st Century" version of the Murphy-Good finite-temperature equation for LECD is

$$J_{\mathrm{L}}^{\mathrm{MGT}}(\phi,T,F) = \lambda_T^{\mathrm{SN}}\;\mathrm{t}_{\mathrm{FD}}^{-2}(f)\;a\phi^{-1}F^2\exp[-\mathrm{v}_{\mathrm{FD}}(f)\cdot b\phi^{3/2}/F]. \tag{E59}$$

where $\lambda_T^{\mathrm{SN}}$ is the SN-barrier form of the temperature correction factor, obtained by putting GPB→SN in eqns (E39) to (E41). The zero-temperature form of this equation is

$$J_{\mathrm{L}}^{\mathrm{MG0}}(\phi,F) = \mathrm{t}_{\mathrm{FD}}^{-2}(f)\;a\phi^{-1}F^2\exp[-\mathrm{v}_{\mathrm{FD}}(f)\cdot b\phi^{3/2}/F]. \tag{E60}$$

## 12. Regime of applicability of the Murphy-Good finite temperature FE equation for LECD

The question arises of the regime of applicability of eq. (E59). Figure S3 is a slightly corrected and reduced version (prepared by Dr M.M. Allaham) of a "regime diagram" originally prepared by Murphy and Good [R14, see their Fig. 6], but with scaled field $f$ plotted on the horizontal axis rather than local "field" $F$.

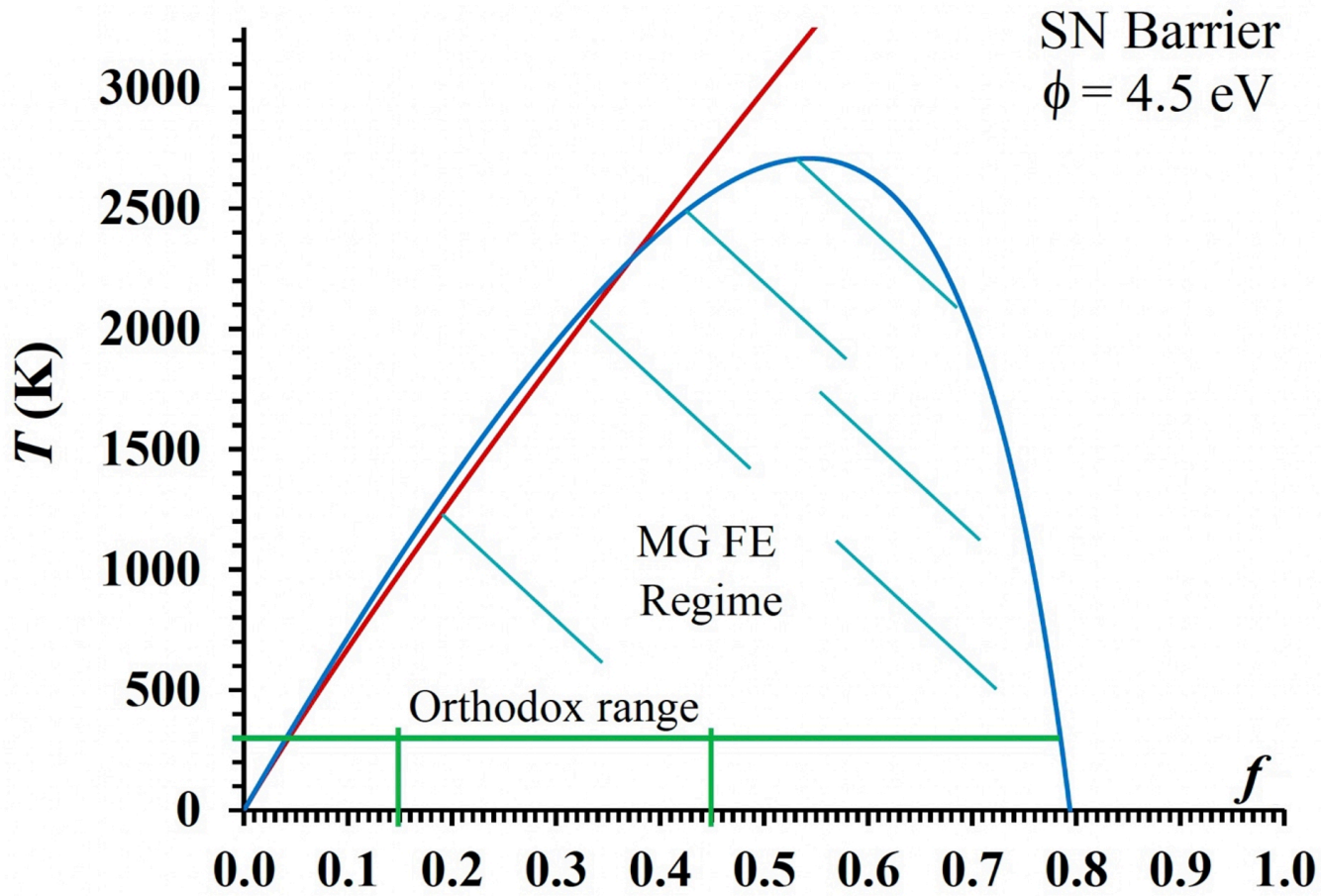

Fig. S3. To illustrate the regime of validity of eq. (E59). According to Murphy and Good [R14], the equation is considered adequately valid within the area inside both the red and blue curves. For a metal with $\phi$= 4.5 eV, $F$ is related to $f$ by $F \approx (14.06 \text{ V/nm})\cdot f$. This version of the diagram, was prepared by Dr M.M. Allaham, and closely resembles the diagram published as Fig. 3b in Ref. [R22].

According to Modinos [R23, discussion following his eq. (1.56)], on the boundary of the emission regime the error in using eq. (E59), as compared with a more exact SPME calculation, varies between 15% and 40% .

The emission regime (i.e. LECD regime) shown in Fig. 3 has been called by various names, including "FE regime", "MG FE regime", "cold field electron emission (CFE) regime" (despite the fact eq. (E59) is valid to quite high temperatures), and "Fowler-Nordheim FE regime". Our view is that possibly the best name would be "Murphy-Good deep-tunnelling FE regime (MG DTFE regime)".

## 13. The Forbes-Deane (FD) special mathematical functions

A special mathematical function (SMF) is a well-defined mathematical function that has well-defined values for every value of its argument. Simple examples are "sin" and "exp". More complicated examples are the Airy functions and the complete elliptical integrals. There exists an international typesetting convention, used here, that symbols for special mathematical function are typeset upright.

The so-called "21st Century" approach to formulating SN barrier theory and applying it to the analysis of FE current-voltage measurements involves both: (a) the function $\mathrm{v}_{\mathrm{FD}}(x)$, introduced above, and (b) a set of related functions, defined below. The whole group of functions (including related functions not defined in this document) can be called the *Forbes-Deane (FD) special mathematical functions*. The SMF $\mathrm{v}_{\mathrm{FD}}(x)$ can be called the *principal Forbes-Deane special mathematical function*, since all the other FD SMFs are derived from $\mathrm{v}_{\mathrm{FD}}(x)$. Since, mathematically, the SMF $\mathrm{v}_{\mathrm{FD}}(x)$ is a special solution of the Hypergeometric Differential Equation first proposed by Euler but extensively investigated by Gauss, it is convenient to call $x$ the *Gauss variable*.

The FD SMFs of interest to the main paper, other than $\mathrm{v}_{\mathrm{FD}}(x)$, are defined as follows. For notational simplicity, here and in some places below, the subscript "FD" is omitted.

$$\mathrm{u}_{\mathrm{FD}}(x) \equiv -\,\mathrm{d}\mathrm{v}/\mathrm{d}x\,, \tag{E61}$$

$$\mathrm{t}_{\mathrm{FD}}(x) \equiv \mathrm{v}_{\mathrm{FD}}(x) - (4/3)\,x\,\mathrm{d}\mathrm{v}/\mathrm{d}x = \mathrm{v}_{\mathrm{FD}}(x) + (4/3)\,x\,\mathrm{u}_{\mathrm{FD}}(x)\,, \tag{E62}$$

$$\mathrm{s}_{\mathrm{FD}}(x) \equiv \mathrm{v}_{\mathrm{FD}}(x) - x\,\mathrm{d}\mathrm{v}/\mathrm{d}x = \mathrm{v}_{\mathrm{FD}}(x) + x\,\mathrm{u}_{\mathrm{FD}}(x)\,. \tag{E63}$$

$$\mathrm{r}_{\mathrm{FD}}(\eta,x) \equiv \exp[\eta\cdot \mathrm{u}_{\mathrm{FD}}(x)]\,. \tag{E64}$$

$\eta$ is considered an abstract mathematical parameter here, but "goes over" to the parameter $\eta$ defined by eq. (E13) when these functions are applied to modelling FE from free-electron metals.

There exist BKH equivalents of these functions. These equivalents use the Nordheim parameter $y$ as the independent variable. These BKH equivalents are in widespread use in the literature, but are not used in the main paper or in this supplementary material.

### 14. Algebraic expressions for $\mathrm{v}_{\mathrm{FD}}(x)$

Because the SMF $\mathrm{v}_{\mathrm{FD}}(x)$ is a special solution of the Gauss HDE. it follows (from the mathematical properties of the Gauss HDE) that there exist many exactly-equivalent exact definitions of this SMF. An exact *analytical* definition does exist (see [R20]), but this involves mathematical functions unfamiliar to most scientists and engineers and is thus of limited use. To generate useful approximations, it is convenient to start from an **exact series expansion** derived in Ref. [R20]. The lowest-order terms of this are

$$\mathrm{v}_{\mathrm{FD}}(x) = 1 - \left(\frac{9}{8}\ln 2 + \frac{3}{16}\right)x - \left(\frac{27}{256}\ln 2 - \frac{51}{1024}\right)x^2 - \left(\frac{315}{8192}\ln 2 - \frac{177}{8192}\right)x^3 + O\left(x^4\right)$$

$$+ x\ln x\left[\frac{3}{16} + \frac{9}{512}x + \frac{105}{16384}x^2 + O\left(x^3\right)\right] \tag{E65}$$

Recurrence formulae for determining higher-order terms are given in [R20].

Note that $\mathrm{v}_{\mathrm{FD}}(x)$ is mathematically unusual, in that (like elliptic integrals) its best series definition needs **two** infinite series.

By dividing the first series in eq. (E65) by (1–$x$), this series definition can be put in the **"second form"**

$$\mathrm{v}_{\mathrm{FD}}(x) = (1-x)\left\{1 + P^{(\infty)}(x)\right\} + (x\ln x)\cdot Q^{(\infty)}(x), \tag{E66}$$

where $P^{(\infty)}(x)$ and $Q^{(\infty)}(x)$ are both infinite power series.

***A high-precision (HP) approximation formula*** for $\mathrm{v}_{\mathrm{FD}}(x)$ is based on the *form* of the series in eq. (E66), but uses a finite number of terms, as below. An HP formula for $\mathrm{u}_{\mathrm{FD}}(x)$, as below, uses a related argument.

$$\mathrm{v}_{\mathrm{HD}}(x) \;\cong\; (1-x)\left[1 + \sum_{i=1}^{4} p_i x^i\right] \;+\; x\ln x \sum_{i=1}^{4} q_i x^{i-1} \tag{E67}$$

$$\mathrm{u}_{\mathrm{HD}}(x) \;\cong\; \mathrm{u}(1) \;-\; (1-x)\sum_{i=0}^{5} s_i x^i \;-\; \ln x \sum_{i=0}^{4} t_i x^i \tag{E68}$$

Best-fit values for the coefficients in these formulae were obtained (by Dr J.H.B. Deane) by standard numerical fitting techniques and are shown in Table S1 below. Over the range $0 \leq x \leq 1$ (but not outside this range), the magnitude of the maximum error in these formulae is $8\times10^{-10}$. [This error is determined by comparison with computer algebra results, which are estimated to have a default accuracy of about 30 decimal places.]

**Table S1.** Constants for use in connection with equations (E67) and (E68).

| $I$ | $p_i$ | $q_i$ | $s_i$ | $t_i$ |
|---|---|---|---|---|
| 0 | - | - | 0.053 249 972 7 | 0.187 5 [=3/16] |
| 1 | 0.032 705 304 46 | 0.187 499 344 1 | 0.024 222 259 59 | 0.035 155 558 74 |
| 2 | 0.009 157 798 739 | 0.017 506 369 47 | 0.015 122 059 58 | 0.019 127 526 80 |
| 3 | 0.002 644 272 807 | 0.005 527 069 444 | 0.007 550 739 834 | 0.011 522 840 09 |
| 4 | 0.000 089 871 738 11 | 0.001 023 904 180 | 0.000 639 172 865 9 | 0.003 624 569 427 |
| $a$ | - | - | –0.000 048 819 745 89 | - |
| $u(1) = u_{HD}(x{=}1) = 3\pi/8\sqrt{2} \cong 0.8330405509$ | | | | |

High-precision formulae for all the Forbes-Deane SMFs, and for various related functions used in FE theory, have been implemented on a downloadable spreadsheet [R24,R25].

***The "simple good approximation" (or "FD06 approximation"),* $\mathbf{v_{FD06}}(x)$.** There also exists a so-called "simple good approximation" for $v_{FD}(x)$, namely

$$v_{FD}(x) \approx v_{FD06}(x) = 1 - x + (1/6)x\ln x \,. \qquad \text{(E69)}$$

This approximation was discovered in 2006 [R18], before the more precise formulae above were developed. This "simple good approximation" is less precise than the formulae set out above, but is simpler to use and seems good enough for most technological purposes.

Over the range $0 \leq x \leq 1$, $v_{FD06}(x)$ has an absolute accuracy better than 0.0024 and relative accuracy better than 0.33%.

***Older approximations.*** In past literature there also exist a large number (about 15) of other relevant approximations. These can be mathematically transformed into approximations for $v_{FD}(x)$ or $v_{FD}(f)$. Most of these approximations have been compared, as approximations for $v_{FD}(f)$, in [R26]. Over the range $0.15 \leq f \leq 0.45$, which is the "pass" range for the orthodoxy test [R27], the simple good approximation is superior to all other published approximations other than a particular form of optimised 3-term approximation due to Forbes and Deane [R19]. The high-precision approximations are significantly superior to all other published approximations.

## 15. Simple good approximations for the Forbes-Deane SMFs

Simple good approximations for the Forbes-Deane SMFs are obtained by applying definitions (E61–64) to eq. (E69). This results in the formulae:

$$v_{FD06}(x) = 1 - x + (1/6)x\ln x \,; \qquad \text{(E70)}$$

$$u_{FD06}(x) = 5/6 - (1/6)\ln x \,; \qquad \text{(E71)}$$

$$t_{FD06}(x) = 1 + (1/9)x - (1/18)x\ln x \,; \qquad \text{(E72)}$$

$$s_{FD06}(x) = 1 - (1/6)x \,; \qquad \text{(E73)}$$

$$r_{FD06}(\eta, x) = \exp[\eta \cdot \{5/6 - (1/6)\ln x\}] \,. \qquad \text{(E74)}$$

These approximations have been used in the main paper.

## 16. "Experiment-oriented" FE equations relating measured current to measured voltage

***Notional emission area for the SN barrier.*** The next stage is to develop (for the SN barrier) an expression for *measured current* $I_m$ in terms of *measured voltage* $V_m$. The first step is to carry out an integration of the LECD over the surface of the emitter. Obviously, real emitters are post-like or needle-like, and surface curvature will in principle affect both the supply of electrons to the emitter surface and the transmission probability. Basic approaches assume the *planar emission approximation*, which postulates that (in the integration) it is a sufficient approximation to take as the LECD at surface location "L" the chosen planar FE equation (here the MG FE equation), using the local work function $\phi_L$ and local surface "field" $F_L$ at location "L". This approximation is adequately valid if the emitter is "not too sharp". Opinions currently differ as to how sharp is "too sharp". Our thinking is that the approximation becomes of deteriorating validity as the apex radius of curvature reduces below about 20–50 nm.

This integration yields a so-called *notional electron-emission current* $I_n^{MGT}$. By choosing a *characteristic location* "C" at which the LECD is particularly high [which implies that ($\phi_C^{3/2}/F_C$) is relatively low], this result can be put in the form

$$I_n^{MGT} = A_{nC}^{SN}\, J_C^{MGT}(\phi_C, T, F_C), \tag{E75}$$

where $A_{nC}^{SN}$ is a *notional emission area*, as derived by assuming a SN barrier, with a particular choice of location "C". In principle, the value of $A_{nC}^{SN}$ depends on the choice of "C" and on various other emitter-related parameters, including emitter shape.

For simplicity in modelling, post and needle-like emitters are usually taken as cylindrically symmetric, and across-surface variations in local work-function are disregarded. (This is the *uniform work-function approximation.*) Hence the subscript on $\phi$ is usually omitted. In these circumstances location "C" is at the emitter apex.

***Characteristic formal emission area for the SN barrier***. $I_n^{MGT}$ is not an accurate prediction of measured emission current $I_m$ because many factors have been left out (in particular, atomic effects). Hence we introduce a *prediction uncertainty factor* $\lambda$ (size unknown) and write

$$I_m = \lambda\, I_n^{MGT} = \lambda\, A_{nC}^{SN}\, J_C^{MGT} \equiv A_{fC}^{SN}\, J_C^{MGT}, \tag{E76}$$

where the *characteristic formal emission area (derived by assuming a SN barrier)*, $A_{fC}^{SN}$, is provisionally given by

$$A_{fC}^{SN} = \lambda\, A_{nC}^{SN}. \tag{E77}$$

However, because the values and functional dependences of $\lambda$ are seriously unknown, this is not a helpful definition for modelling or for data analysis. It is better to *define* the formal emission area $A_{fC}^{SN}$ in the following way. First, the *kernel current density* $J_k^{MG}$ *for the MG FE equation* is defined by omitting the $t_{FD}^{-2}(f)$ term from eq. (E60) to give

$$J_k^{MG}(\phi, F) = a\phi^{-1}F^2\exp[-v_{FD}(f)\cdot b\phi^{3/2}/F]. \tag{E78}$$

The formal emission area $A_{fC}^{SN}$ is then regarded a parameter *derived from experiment*, given by

$$A_{fC}^{SN} \equiv I_m/J_{kC}^{MG}\ , \tag{E79}$$

***Voltage-to-local-field conversion factor (CF).*** The next step is to define the relationship between measured voltage $V_m$ and the characteristic local field $F_C$ (equivalent, in models with cylindrical symmetry, to apex local field). This relationship is written here in the form

$$F_C = \beta_C V_m\ , \tag{E80}$$

where $\beta_C$ is the *characteristic (voltage-to-local-field) conversion factor (CF)* defined via this equation

***EMG equation for $I_m(V_m)$.*** The equation for $I_m$ as a function of $V_m$ can then be written in the form

$$I_m^{EMG} = A_{fC}^{SN}\, a\phi^{-1}(\beta_C V_m)^2 \exp[-v_{FD}(f_C)\, b\phi^{3/2}/\, \beta_C V_m]\ . \tag{E81}$$

This equation has been called the *"extended" or "experiment-oriented" Murphy-Good (EMG) FE equation* (for measured current as a function of measured voltage).

The effect of this "experiment-oriented" approach has been to "sweep into the formal area" the small correction terms associated with temperature and with the Taylor expansion of $G^{\mathrm{SN}}$, where they join the much larger correction terms hidden in the formal emission area, in particular those associated with the neglect of effects due to atomic wave-functions. Thus the interpretation of measured current-voltage characteristics is split into two tasks: the accurate extraction of formal area values from experiments; and the physical interpretation of these values. Due to gaps in our understanding of quantum mechanics and/or how best to apply it, amongst other things, full interpretation of these values is not possible at present, but $A_{\mathrm{fC}}^{\mathrm{SN}}$ could be a a useful empirical parameter in research-and-development contexts.

***Characteristic formal emission area for the ET barrier.*** If an exactly triangular (ET) barrier is assumed, rather than a SN barrier, then an argument similar to that above leads to the following definition for the *characteristic formal emission area (derived by assuming an ET barrier)*, $A_{\mathrm{fC}}^{\mathrm{ET}}$:

$$A_{\mathrm{fC}}^{\mathrm{ET}} \equiv I_{\mathrm{m}} / J_{\mathrm{C}}^{\mathrm{EL}}, \tag{E82}$$

where $J_{\mathrm{C}}^{\mathrm{EL}}$ is the characteristic LECD given by the elementary FE equation, with (as before) the characteristic location being taken at the emitter apex in the simplest emitter models.

## 17. Analysis of Fowler-Nordheim plots using the Extended Murphy-Good FE equation

An FE system in which there is no leakage current and the conversion factor $\beta_{\mathrm{C}}$ is constant, independent of measured voltage $V_{\mathrm{m}}$, is said to be *electronically ideal*. Conventional methods of FE data-analysis, including Fowler-Nordheim (FN) plots, may yield spurious results if the FE system and the resulting data are not electronically ideal. The need for validity tests is discussed below. The data analysis method discussed in this section assumes that the FE system that generated the data is electronically ideal.

It is best practice to make experimental Fowler-Nordheim plots by using the measured current-voltage data, and this is assumed here.

In so-called *Fowler-Nordheim (FN) coordinates*, eq. (E81) takes the form

$$L_{\mathrm{FN}}^{\mathrm{EMG}} \equiv \ln\{I_{\mathrm{m}}^{\mathrm{EMG}}/V_{\mathrm{m}}^{2}\} = \ln\{A_{\mathrm{fC}}^{\mathrm{SN}}\, a\phi^{-1}\beta^{2}\} - \mathrm{v}_{\mathrm{FD}}(f_{\mathrm{C}})\, b\phi^{3/2}/\, \beta V_{\mathrm{m}} \tag{E83}$$

This equation describes a slightly curved line (more so at the high-voltage end).

The *tangent method* of FN plot analysis was introduced in 1953 by Burgess, Kroemer and Houston [R15]. A modified version of the original method is used here. In this approach, the line fitted to the experimental data points is modelled as a tangent to eq. (E83), taken at the (unknown) voltage value $V_{\mathrm{t}}$ at which this tangent is parallel to the straight line fitted to the experimental points.

Strictly, the straight line fitted to the experimental points should be regarded as a chord to eq. (E83) and a small correction should be applied to the tangent method (see [R28]). However, the size of the correction is small and it is almost always disregarded.

This value $V_{\mathrm{t}}$ is termed the *fitting value* of $V_{\mathrm{m}}$. As shown below, $V_{\mathrm{t}}$ corresponds to a particular (but initially unknown) fitting value $f_{\mathrm{t}}$ of characteristic scaled field $f_{\mathrm{C}}$.

It follows that, at the fitting point, the tangent to eq. (E83) can be written in the form

$$L^{\tan} = \ln\{R^{\tan}(f_{\mathrm{t}})\} + S^{\tan}(f_{\mathrm{t}}) / V_{\mathrm{m}} . \tag{E84}$$

It can be shown (see Appendix A2) that the slope $S^{\tan}(f_{\mathrm{t}})$ of this tangent is given by

$$S^{\tan}(f_{\mathrm{t}}) = -\, \mathrm{s}_{\mathrm{FD}}(f_{\mathrm{t}}) \cdot b\phi^{3/2}/\beta_{\mathrm{C}} , \tag{E85}$$

where the fitting value $\mathrm{s}_{\mathrm{FD}}(f_{\mathrm{t}})$ of the *slope correction factor* $\mathrm{s}_{\mathrm{FD}}(f_{\mathrm{C}})$ is a particular value of the SMF $\mathrm{s}_{\mathrm{FD}}(x)$ defined earlier.

The parameter $\ln\{R^{\tan}(f_t)\}$ gives the value of $L_{FN}$ at which the tangent intersects the vertical ($1/V_m$=0) axis. It can be shown [R29] that $R^{\tan}(f_t)$ can be written in the form

$$R^{\tan}(f_t) \;=\; \mathrm{r}_{FD}(\phi,f_t)\, A_f^{SN}\, a\phi^{-1}\beta_C^{\,2}\,, \tag{E86}$$

where the fitting value $\mathrm{r}_{FD}(\phi,f_t)$ of the *intercept correction factor* $\mathrm{r}_{FD}(\phi,f_C)$ is a particular value of the SMF $\mathrm{r}_{RD}(\eta,x)$ defined earlier.

For notational simplicity, define $\mathrm{r}_t \equiv \mathrm{r}_{RD}(\phi,f_t)$ and $\mathrm{s}_t \equiv \mathrm{s}_{FD}(f_t)$. Combining (E85) and (E86) then yields

$$R^{\tan}(f_t)\cdot\{S^{\tan}(f_t)\}^2 \;=\; A_f^{SN}\cdot(ab\phi^2)\cdot(r_t s_t^{\,2}). \tag{E87}$$

Let the straight line fitted to the experimental FN plot have the form

$$L^{expt} \;=\; \ln\{R^{expt}\} + S^{expt}/V_m\,. \tag{E88}$$

By identifying the experimental slope and intercept with the slope and intercept of the theoretical tangent, the following extraction formulae are obtained:

$$\beta^{extr}[\mathrm{SN}] \;=\; -\,\mathrm{s}_t\, b\phi^{3/2}\,/\,S^{expt}\,; \tag{E89}$$

$$\{A_f^{SN}\}^{extr} \;=\; \Lambda_{FN}^{SN}\cdot[R^{expt}\,(S^{expt})^2], \tag{E90}$$

where the *area extraction parameter*, $\Lambda_{FN}^{SN}$, for a FN plot, assuming a SN barrier, is given by

$$\Lambda_{FN}^{SN} = 1/[(ab^2\phi^2)\cdot(\mathrm{r}_t\mathrm{s}_t^{\,2})]. \tag{E91}$$

These formulae apply to both single-tip field electron emitters (STFEs) and to large-area field electron emitters (LAFEs). For LAFEs, an additional step is possible, in that one can derive an extracted value for the *formal area efficiency (assuming an SN barrier)*, $\alpha_f^{SN}$. If the *"macroscopic area" (or "footprint")* $A_M$ has been recorded, then this extracted value is given by:

$$\{\alpha_f^{SN}\}^{extr} \;=\; \{A_f^{SN}\}^{extr}/A_M. \tag{E92}$$

This parameter $\{\alpha_f^{SN}\}^{extr}$ is a measure of the fraction of the emitter's macroscopic area that is emitting electrons. It is not an accurate measure, but is a characterisation parameter that could be useful in Research-and-Development contexts.

## 18. Evaluation of the fitting parameters

It remains to determine suitable values for the various fitting parameters. The relevant equations are as follows. From definition (E20), with $f_\phi$ set equal to $f_t$, and definition (E80), with all parameters subscripted "t", we obtain

$$f_t \;=\; (c^2\phi^{-2}\,V_t)\cdot\beta_t\,. \tag{E93}$$

From eq. (E89) we have

$$\beta_t = \;-\,\mathrm{s}_t\, b\phi^{3/2}\,/\,S^{expt}\,. \tag{E94}$$

with $\mathrm{s}_t$ estimated, as earlier, via the relation

$$\mathrm{s}_t \;\equiv\; \mathrm{s}_{FD06}(x{=}f_t)\,. \tag{E95}$$

It can be seen that in eq. (E93) there are two unknowns, $V_t$ and $\beta_t$. In older work, normal practice was to choose a specific value for $\mathrm{s}_t$ (often 0.95) from which everything else can be determined. In the present work a different approach is taken, namely a specific value is chosen for $V_t$, a starting value $\beta_{start}$ is chosen for the conversion factor, and an iterative process is used to determine good-precision values for $f_t$, $\mathrm{s}_t$ and other relevant parameters.

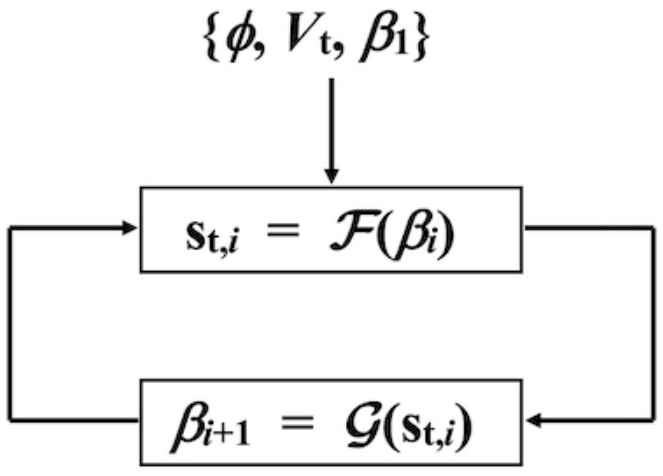

Fig. S4. To illustrate the iteration loop in the authors' data-analysis program.

The iteration loop is illustrated in Fig. S4 and involves the two components

$$\mathrm{s}_{\mathrm{t},i} \;=\; \mathcal{F}(\beta_i) \;=\; \mathrm{s}_{\mathrm{FD06}}(f_{\mathrm{t},i}),\ \text{with}\ f_{\mathrm{t},i} \;=\; \{c^2\phi^{-2}\,V_{\mathrm{t}}\}\cdot\beta_i\,, \tag{E96}$$

$$\beta_{i+1} \;=\; \mathcal{G}(\mathrm{s}_{\mathrm{t},i}) \;=\; -\,\mathrm{s}_{\mathrm{t},i}\cdot\{b\phi^{3/2}/S^{\mathrm{expt}}\}. \tag{E97}$$

Here, $i$ is the iteration variable ($i \geq 1$), $\beta_i$, $f_{\mathrm{t},i}$ and $\mathrm{s}_{\mathrm{t},i}$ are the $i$th estimates of $\beta$, $f_{\mathrm{t}}$ and $\mathrm{s}_{\mathrm{t}}$, and $\mathcal{F}(\beta_i)$ and $\mathcal{G}(\mathrm{s}_{\mathrm{t},i})$ are the modelling functions shown. For notational simplicity here, all labels other than "t" and "$i$" have been removed.

We choose some particular value in the measured voltage range to be an estimate of the fitting value $V_{\mathrm{t}}$. To start the iteration we set $\beta_1$ equal to a chosen *starting-value* $\beta_{\mathrm{start}}$. In practice, it is found sufficient to take $\beta_{\mathrm{start}} = 10^{-3}\ \mathrm{nm}^{-1}$, The iteration rapidly converges, and is stopped when the incremental change in $\beta$ is less than $10^{-4}$. The resulting estimates of the fitting parameters can then be used to extract characterization parameters.

The process is then repeated for a set of values of $V_{\mathrm{t}}$ chosen systematically from across the measured-voltage range.

The output of the program shows how the various calculated parameters depend on the value assumed for the fitting voltage. The further analysis of this data is described in Section 2.3 of the main paper.

## 19. Overview of downloadable program

A program has been developed that implements this iteration loop in the context of using a FN plot to analyse current-voltage-type input data. The program then goes on to extract emitter characterization parameters. A separate part of the program is available to calculate the FE total-energy distribution, as a function of work function, temperature and field. The high-level structure of the main program option is as shown in Table S2.

Table S2. Effective high-level structure of main programme option

| 1. | Select assumed value of local work function $\phi$ |
|---|---|
| 2. | Read in experimental $I_{\mathrm{m}}(V_{\mathrm{m}})$ data |
| 4. | Fit regression line to FN plot of input data, and extract values of $S^{\mathrm{expt}}$ and $R^{\mathrm{expt}}$ |
| 5. | Use range of measured-voltage values to decide estimates of fitting voltage $V_{\mathrm{t}}$ to be assessed. |
| 6. | For each chosen fitting-value estimate $V_{\mathrm{t}}$ , run steps 7 to 10 |
| 7. | Run iteration loop to find related fitting values of $f_{\mathrm{t}}$, $\mathrm{s}_{\mathrm{t}}$ and $\mathrm{r}_{\mathrm{t}}$ and extracted value of $\beta$ |
| 8. | Extract related value of formal emission area $A_{\mathrm{f}}^{\mathrm{SN}}$ |
| 9. | Calculate, for chosen value of $V_{\mathrm{t}}$ and related extracted value of $\beta$, the related extracted value of field $F$ and the related extracted value of decay width $d_{\mathrm{F}}$ |
| 10. | Calculate corresponding value of practical brightness, using eq. (9) in main paper |
| 11. | Report results in graphical form as functions of chosen fitting voltage |

## 20. Interpretation of formal emission area $A_\mathrm{f}^\mathrm{SN}$

It needs to be understood that the formal emission area $A_\mathrm{f}^\mathrm{SN}$ is **not** an accurate estimate of "true emission area" (however this is defined), but is a "catch all" parameter that **also** includes effects due to the following: limitations of Murphy-Good theory, including all quantum-mechanical limitations, especially neglect of effects due to atomic-level wave-functions and neglect of effects due to emitter electronic band structure; temperature effects; effects related to the first-order Taylor expansion of $G^\mathrm{SN}$, effects due to work-function variability across the surface; emitter shape effects of various kinds; and accuracy-limitations of FN-plot analysis.

It will probably be many years (possibly 20 years or more) before we achieve an understanding of FE theory that is good enough to allow accurate extraction of "true emission area" (however this is defined) from experimental data by using FN-plot or similar techniques.

Notwithstanding this, both $\{A_\mathrm{f}^\mathrm{SN}\}^\mathrm{extr}$ and $\{\alpha_\mathrm{f}^\mathrm{SN}\}^\mathrm{extr}$ are considered to be useful empirical parameters that (particularly in the context of Research and Development) can be used to compare the emission properties of different emitters.

Slightly better precision in extraction of formal emission area $A_\mathrm{f}^\mathrm{SN}$ can in principle be obtained (for electronically ideal systems) if a so-called Murphy-Good plot is used, rather than a FN plot. Details are given in [R30]. This approach was considered beyond the scope of the present work, but could usefully be investigated in future work.

## 21. Extraction of parameters from FN plots using exactly-triangular-barrier theory

As above, the remarks in this Section apply only to FE systems and data that are electronically ideal.

Arguments broadly similar to those above, but somewhat simpler, can be applied to the *exactly triangular (ET) barrier*. These result in the following extraction equations for characterization parameters:

$$\beta^\mathrm{extr}[\mathrm{ET}] = -b\phi^{3/2} / S^\mathrm{expt}, \tag{E98}$$

$$\{A_\mathrm{f}^\mathrm{ET}\}^\mathrm{extr} = \Lambda_\mathrm{FN}^\mathrm{ET} \cdot [R^\mathrm{expt}\,(S^\mathrm{expt})^2], \tag{E99}$$

where the *area extraction parameter*, $\Lambda_\mathrm{FN}^\mathrm{ET}$, for a FN plot, assuming an ET barrier, is given by

$$\Lambda_\mathrm{FN}^\mathrm{ET} = 1/[(ab^2\phi^2)]. \tag{E100}$$

As can be seen, the results for the ET barrier are obtained from those for the SN barrier by setting $s_\mathrm{t}$=1 and $r_\mathrm{t}$=1. For the conversion factor the difference is typically about 5%. But for formal emission areas, the ET-barrier approach yields formal emission areas that are typically about a factor of 100 (or more) greater than those derived used the SN-barrier approach. Thus, theoretical expectation is that significant error is involved in emission-area extraction when the ET-barrier approach is used.

## 22. The need for validity checks

### *22.1 The causes of non-ideality*

As already indicated, the data-analysis methods discussed here will yields correct results only if the FE system and associated current-voltage data are "electronically ideal". Aspects of this concept can be understood by reference to the idealised circuit diagram shown as Fig. S5.

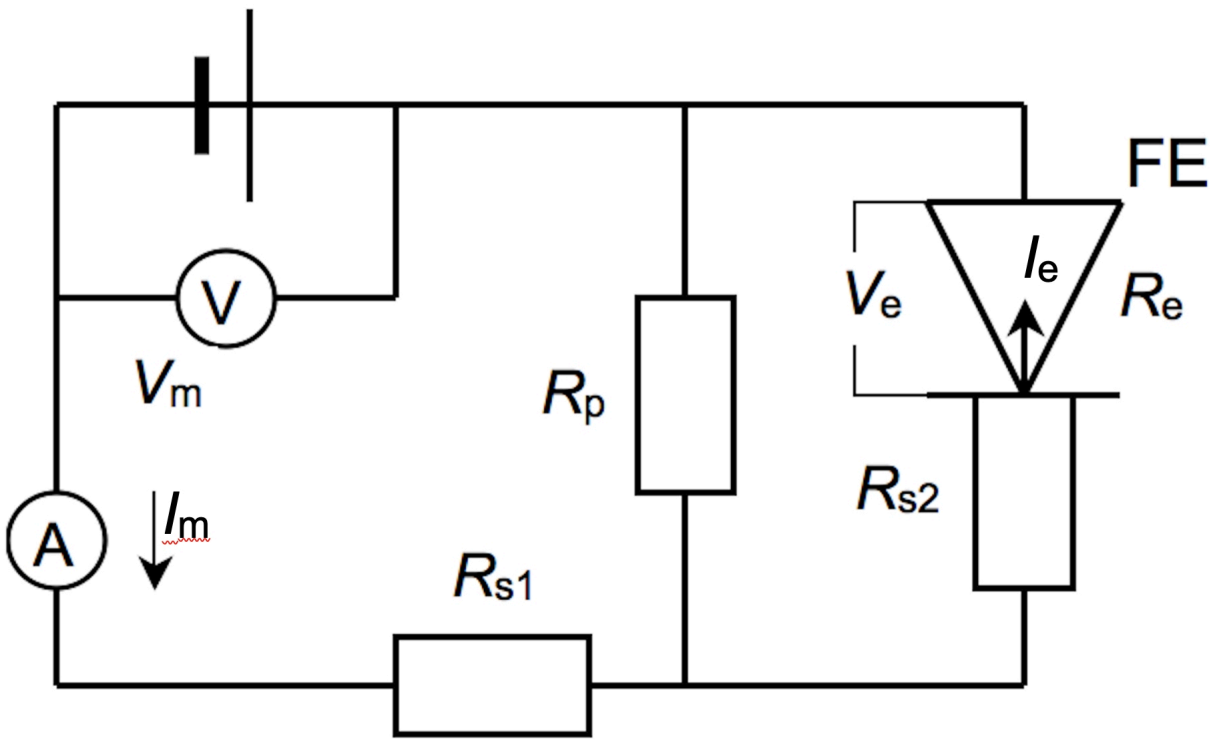

Fig. S5 Idealised circuit diagram of a FE system. For precise definitions of the emission voltage $V_e$ and emission current $I_e$ see text. Note that this diagram uses the electron emission convention in which electron currents are treated as positive, even though they would be negative in conventional electricity theory.

In this circuit diagram, the *emission voltage* $V_e$ is the voltage between the counter-electrode and the characteristic location "C" on the surface of the emitter apex region. $V_e$ is a positive quantity in the circumstances of field electron emission. The *emission current* $I_e$ is the electron-current that passes from the emitter apex region to the counter-electrode. The emission resistance $R_e \equiv V_e/I_e$, and is a function of $V_e$ (and also of the effective area of emission). $R_e$ is high for low values of $V_e$, but steadily decreases as $V_e$ increases.

A FE system is called *electronically ideal* if: (1) $I_e=I_m$; and (2) $V_e=V_m$. The first condition requires that the *parallel resistance* $R_p$ be very large (in comparison with the ($R_{s2}+R_e$), so that there is no l*eakage current* through $R_p$. The second condition requires that the total *series resistance* ($R_{s1}+R_{s2}$) be very small in comparison with $R_e$. The first condition can often be implemented by suitable system design. The second condition can sometimes be difficult to achieve, particularly with non-metallic emitters that have relatively high resistivity.

Because the value of $R_e$ falls off with increasing $V_e$, it is possible for a FE system to be electronically ideal in the lower part of its measured operating voltage ($V_m$) range, but to become non-ideal at higher operating voltages, typically causing a Fowler-Nordheim plot made using measured data to "bend downwards" at the left-hand (high-voltage) side.

There are also other situations, such as the presence of field emitted vacuum space-charge (FEVSC), that can cause non-ideality. A more comprehensive discussion, identifying 12 forms of "system complication", i.e. 12 potential causes of non-ideality, is provided in Ref. [R31]. However, it is thought by one of us (RGF) that the commonest (though not the only) cause of observed non-ideality is probably voltage loss along the shank of the emitter. Many papers seem not to recognise this as a possibility.

There are thought to be many published FE papers that state spurious values of characterization parameters because their original data are not electronically ideal. Hence, *before* carrying out detailed data analysis it is advisable to check that the FE system being used is electronically ideal (or nearly so), at least over a significant part of the measured voltage range. It is also useful to report in presentations and papers that this has been done.

Relevant validity checks are: (1) that the chosen data-analysis plot is "nearly linear"; and (2) that the system passes the Orthodoxy Test [R27]. For detailed discussion see [R32]. For a spreadsheet for applying the Orthodoxy Test, see supplementary material for [R27]. A webtool for applying the Test is under development [R22,R33].

If the Orthodoxy Test has not been carried out before detailed data analysis, there is now a simplified version of the test (provisionally called the "magic emitter test" [R31]), but likely to end up as called the "implausible field-values test") that can be applied after data analysis, if information is available both about the highest macroscopic-field value used in the experiments and about an extracted dimensionless field enhancement factor (FEF). This "magic emitter" test may also be useful when reviewing papers or testing the validity of published FEF values. A merit of all the tests is that they can easily be applied by graduate students without needing assistance from their supervisors.

### Appendix A1: Derivation of expression for Second FN Constant $b$

The simplest method of deriving an expression for the Second FN Constant is to apply the sJWKB formalism to the ET barrier. From eq. (E15), the motive energy for an ET barrier of zero-field height $H$, in the presence of local "field" $F$, is

$$M^{\mathrm{ET}}(H,F,z) \equiv H - eFz \,. \qquad \text{(A1.1)}$$

The range of integration needed (where the motive energy is $\geq 0$) is from $z=0$ to $z=H/eF$. Hence, from eq. (E23), the Gamow factor for the ET barrier is

$$G^{\mathrm{ET}} = 2\kappa_{\mathrm{e}} \int_0^{H/eF} (H - eFz)^{1/2} \mathrm{d}z \; = \; -(4/3)(\kappa_{\mathrm{e}}/eF)\left[(H - eFz)^{3/2}\right]_0^{H/eF}$$
$$= \; (4/3)(\kappa_{\mathrm{e}}/e)H^{3/2}/F. \qquad \text{(A1.2)}$$

From eq. (E1), $\kappa_{\mathrm{e}} \equiv (2m_{\mathrm{e}})^{1/2}/\hbar$. It follows that expression (A1.2) can be rewritten in the form

$$G^{\mathrm{ET}} \; = \; bH^{3/2}/F, \qquad \text{(A1.3)}$$

where $b$ is an universal constant defined by

$$b \equiv (4/3)(2m_{\mathrm{e}})^{1/2}/e\hbar \,. \qquad \text{(A1.4)}$$

This constant, $b$, also emerges in more complicated treatments, either of the ET barrier or of approximately triangular barriers.

### Appendix A2: Derivation of expressions for slope correction function $\mathrm{s}_{\mathrm{FD}}(f_{\mathrm{C}})$ and for SMF $\mathrm{s}_{\mathrm{FD}}(x)$

From eq. (E82), when the (theoretical) EMG FE equation is applied to a needle or post-like emitter and is plotted in $I_{\mathrm{m}}(V_{\mathrm{m}})$-type FN coordinates, then it takes the form

$$L_{\mathrm{FN}}^{\mathrm{EMG}} \equiv \ln\{I_{\mathrm{m}}^{\mathrm{EMG}}/V_{\mathrm{m}}^{2}\} \; = \; \ln\{A_{\mathrm{fC}}^{\mathrm{SN}}\, a\phi^{-1}\beta_{\mathrm{C}}^{2}\} - \mathrm{v}_{\mathrm{FD}}(f_{\mathrm{C}})\, b\phi^{3/2}/\, \beta_{\mathrm{C}}V_{\mathrm{m}} \,, \qquad \text{(A2.1)}$$

This equation describes a slightly curved line (more so at the high-voltage end).

If any weak field dependences in $\phi$ or in $A_{\mathrm{fC}}^{\mathrm{EMG}}$ are disregarded, then—at any specific value of $V_{\mathrm{m}}^{-1}$ (and hence some specific value of $f_{\mathrm{C}}$)—the slope $S^{\mathrm{theor}}(V_{\mathrm{m}}^{-1})$ of eq. (A2.1)—and hence the slope of the tangent to this curve, taken at this value of $V_{\mathrm{m}}^{-1}$, is given by

$$S^{\tan}(V_{\mathrm{m}}^{-1}) \equiv \mathrm{d}L_{\mathrm{FN}}/\mathrm{d}(V_{\mathrm{m}}^{-1}) = -\left(\partial\{\mathrm{v}_{\mathrm{FD}}(f_{\mathrm{C}})\cdot V_{\mathrm{m}}^{-1}\}/\partial\{V_{\mathrm{m}}^{-1}\}\right)_{\phi} \cdot b\phi^{3/2}/\beta_{\mathrm{C}}$$
$$= -\left[\, \mathrm{v}_{\mathrm{FD}}(f_{\mathrm{C}}) + V_{\mathrm{m}}^{-1} \left(\partial\{\mathrm{v}_{\mathrm{FD}}(f_{\mathrm{C}})\}/\partial\{V_{\mathrm{m}}^{-1}\}\right)_{\phi}\right] \cdot b\phi^{3/2}/\beta_{\mathrm{C}}. \qquad \text{(A2.2)}$$

where the fact that $\mathrm{v}_{\mathrm{FD}}(f_{\mathrm{C}})$ is an indirect function of measured voltage has been taken into account. Further, since $\mathrm{d}(V_{\mathrm{m}}^{-1}) \; = \; -V_{\mathrm{m}}^{-2}\mathrm{d}V_{\mathrm{m}}$, eq. (A2.2) becomes

$$S^{\tan}(V_{\mathrm{m}}^{-1}) \; = \; -\left[\mathrm{v}_{\mathrm{FD}}(f_{\mathrm{C}}) - V_{\mathrm{m}} \left(\partial\{\mathrm{v}_{\mathrm{FD}}(f_{\mathrm{C}})/\partial V_{\mathrm{m}}\right)_{\phi}\right] \cdot b\phi^{3/2}/\beta_{\mathrm{C}}. \qquad \text{(A2.3)}$$

From the relations $f_{\mathrm{C}}=F_{\mathrm{C}}/F_{\phi}$ and $F_{\phi}=\beta_{\mathrm{C}}V_{\mathrm{m},\phi}$ it is possible to define a *reference measured voltage* $V_{\mathrm{m},\phi}$ at which a SN barrier of zero-field height $\phi$ is reduced to zero, and deduce that:

$$V_{\mathrm{m}} \; = \; f_{\mathrm{C}}\, V_{\mathrm{m},\phi} \,; \;\text{ and }\; \mathrm{d}V_{\mathrm{m}} \; = \; V_{\mathrm{m},\phi}\, \mathrm{d}f_{\mathrm{C}} \,. \qquad \text{(A2.4)}$$

Hence:

$$V_{\mathrm{m}}/\mathrm{d}V_{\mathrm{m}} \; = \; f_{\mathrm{C}}/\mathrm{d}f_{\mathrm{C}} \,, \qquad \text{(A2.5)}$$

and eq, (A2.3) can be rewritten as

$$S^{\tan}(V_{\mathrm{m}}^{-1}) = -[\mathrm{v}_{\mathrm{FD}}(f_{\mathrm{C}}) - f_{\mathrm{C}}\,(\partial\{\mathrm{v}_{\mathrm{FD}}(f_{\mathrm{C}})/\partial f_{\mathrm{C}})_{\phi}] \cdot b\phi^{3/2}/\beta_{\mathrm{C}}\,. \quad \text{(A2.6)}$$

The expression in square brackets can be identified as a particular application of the SMF $\mathrm{s}_{\mathrm{FD}}(x)$ defined by eq. (E63), and hence can be represented by the symbol $\mathrm{s}_{\mathrm{FD}}(f_{\mathrm{C}})$. Thus eq. (A2.6) becomes written

$$S^{\tan}(V_{\mathrm{m}}^{-1}) = -\,\mathrm{s}_{\mathrm{FD}}(f_{\mathrm{C}}) \cdot b\phi^{3/2}/\beta_{\mathrm{C}}. \quad \text{(A2.7)}$$